\documentclass[12pt,a4paper]{article}
\usepackage{amssymb,amsmath,amscd,epsfig}
\usepackage[dvips]{color}
\usepackage{bm}
\newcommand{\be}{\begin{eqnarray}}
\newcommand{\ee}{\end{eqnarray}}
\newcommand{\bi}{\bibitem}

\newcommand{\rar}{\rightarrow}
\newcommand{\lrar}{\leftrightarrow}

\newcommand{\mnu}{m_\nu}

\newcommand{\dm}{\delta m^2}

\newcommand{\tcred}{\textcolor{red}}
\newcommand{\vs}{\vspace}

\definecolor{gold}{rgb}{0.89,0.78,0}
\definecolor{grn05}{rgb}{0,0.5,0}

\newcommand{\nue}{\nu_e} 
\newcommand{\num}{\nu_\mu}
\newcommand{\nut}{\nu_\tau}

\begin{document}
\title{{
{
{\bf INTRODUCTION TO COSMOLOGY}
}}}
\author{
{{{A.D. Dolgov}}}
\\[5mm]
\it {{ ITEP, 117218, Moscow, Russia}}\\
{\it { INFN, Ferrara 40100, Italy}}\\
{\it { University of Ferrara, Ferrara 40100, Italy}}
}
\vspace{2cm}
\date{{ {{
ITEP Winter School}}\\
{{ Moscow
}}}
{{
\begin{center}{
February 9-14, 2009
}\end{center}
}
}}
\maketitle
\vs{0.4cm}

\begin{abstract}
An introductory lectures on cosmology at ITEP Winter School
for students specializing in particle physics are presented. Many important 
subjects are not covered because of lack of time and space but hopefully the
lectures may serve as a starting point for further studies.  
\end{abstract}

\section{Introduction \label{s-intro}}

Modern cosmology is vast interdisciplinary science and it is
impossible to cover it in any considerable detail in five hours
allocated to me at this School. The task is even more difficult
because of different background and level of the participants.
Planning these lectures, I have prepared the following short list of
subjects, which is surely will be made much shorter at this lecture course,
but hopefully it may be useful for the students who would like to continue
studying this field. So the idealistic content could be the following:
\begin{enumerate}
\item{{A little about general relativity and its role in cosmology.}}
\item{{ Four basic cosmological equations and expansion regimes.}}
\item{ Universe today and in the past.} 
\item{{Kinetics in hot expanding world and
freezing of species.}}
\item{{Inflation:} kinematics, models, universe heating, and 
generation of density perturbations and gravitational waves.}
\item{{Big bang nucleosynthesis.}}
\item{{Field theory at non-zero temperature and
cosmological phase transitions.}}
\item{{Baryogenesis and cosmological antimatter.}} 
\item{{Neutrino in cosmology (bounds on mass, oscillations,
magnetic moment, and anomalous interactions.}}
\item{Dark matter and large scale structure (LSS).}
\item{{Vacuum and dark energies.}} 
\item{{Cosmic microwave radiation (CMB) and cosmological parameters.}}
\end{enumerate}
In reality about a half of this plan was fulfilled. At least this
lectures could be helpful for a first aquaintance with cosmology and as
starting point for deeper studies. 

We will start from some non-technical introduction to General Relativity
and relations between the latter and cosmology, sec.~\ref{s-grav-cosm}.
Next we will derive the basic cosmological equations ina rather naive
way studying motion of non-relativistic test body in spherically symmetric
gravitational field, sec.~\ref{s-expan}. There we also talk about
realistic regimes of the universe expansion and basic cosmological paramters.
In the next section, \ref{s-history} the universe history is very briefly
presented. Section \ref{s-hot} is dedicated to thermodynamics and kinetics
in the early universe.  Section \ref{s-freeze} is dedicated to freezing of
species and cosmological limit on neutrino mass. Big bang nucleosynthesis is
presented in sec.~\ref{s-bbn}. In section~\ref{s-nu-bbn} the role of neutrinos
in BBN is described. Neutrino oscillations in the early universe are considered
in sec.~\ref{s-nu-osc}. In section~\ref{s-infl} inlationary cosmology is discussed, and
the last section~\ref{s-bg} is dedicated to cosmological baryogenesis.

\section{Gravity and cosmology \label{s-grav-cosm}}

Two simple observations that the sky is dark at night and that there are
shining stars lead to the conclusion that the universe is finite in space and
time. The first one is the well known Olbers' paradox, based on the estimate
of the sky luminosity, which in infinite homogeneous static universe must be
infinitely high. Shining stars should exhaust their fuel in finite time and 
thus cannot exist in the infinitely old universe -- thermal death of the 
universe. General relativity (GR) successfully hit both targets leading
to the notion of expanding universe of finite age, but created instead
its own very interesting problems which we discuss in what follows.

Newtonian theory of gravity has an evident shortcoming that it
has action-at-a-distance property. In other words,
gravitation acts instantaneously, at any distance. 
On the other hand, in the spirit
of contemporary wisdom interactions are always mediated by some bosonic
fields and are relativistically invariant. If we wished today to 
generalise Newtonian theory of gravity to relativistic theory
we could take, {\it a priori} as a mediator of interactions scalar, 
vector, or tensor intermediate bosons, confining ourselves to lower spins.

Since we know that gravity operates at astronomically large distances,
the mass of the intermediate boson should be zero or very small. Indeed,
massless bosons create static Coulomb type potential, $U\sim 1/r$,
while massive bosons lead to exponentially cut-off Yukawa potential,
$U \sim \exp(-mr)/r$.

Interactions mediated by vector field are odd with respect to charge parity
transformation, C-transformation, and as one can see from the 
vector boson propagator, such interactions induce matter-antimatter
attraction and matter-matter repulsion, recall electromagnetic interactions. 
Hence vector field cannot mediate attractive gravitational force.

Scalar and tensor mediators lead to attraction of matter-matter
and matter-antimatter and both are {\it a priori} allowed.
According to non-relativistic Newtonian theory the 
source of gravity is mass. Possible relativistic generalisation 
for scalars should be a scalar quantity coinciding in non-relativistic 
limit with mass. The only known such source is 
the trace of the energy-momentum tensor
of matter, ${ {T_{\mu}^{\mu}}}$. The relativistic equation of motion for scalar
gravity should have the form:
\be{{
\partial^2 \Phi = 8\pi G_N T_{\mu}^{\mu},
}}\label{scalar-eq}
\ee
where $G_N$ is the Newtonian gravitational coupling constant.
Such theory is rejected by the observed 
light bending in gravitational field, since for photons:
${{{ T_\mu^\mu = 0}.}}$
A small admixture of scalar gravity to tensor one, i.e.
Brans-Dicke theory~\cite{brans}, is allowed.

There remains massless tensor theory with the source which may be only
the energy-momentum tensor of matter, ${{ T_{\mu\nu}}}$. 
In first approximation the equation of motion takes the form:
\be{{{
\partial^2 h_{\mu\nu} = 8\pi G_N T_{\mu\nu}.
}}}\label{tensor-eq1}
\ee
This equation is valid in the weak field approximation because the 
energy-momentum of ${{ h_{\mu\nu}}}$ itself should be included to ensure
conservation of the {\it total} enegy-momentum.

Massless particles, as e.g. gravitons, must interact with a conserved 
source. Otherwise theory becomes infrared pathological. The energy-momentum
tensor of matter is conserved only if the energy transfer to gravitational
field is neglected. Taking into account energy leak into gravity leads
to non-linear equations of motion and
allows to reconstruct GR order by order. For a discussion of this 
approach see papers~\cite{lpg}. 

Historically Einstein did not start from field theoretical approach but
formulated general relativity in an elegant and economical way 
as geometrical theory postulating that
matter makes space-time curved and that the motion of matter in
gravitational field is simply free fall along geodesics of this curved
manifold. This construction is heavily based on the universality of
gravitational action on all types of matter -- the famous equivalence
principle, probably first formulated by Galileo Galilei. 
The least action principle for GR was formulated by Hilbert with the
action given by
\be 
A = \frac{ 1}{16\pi G_N} 
\int d^4 x \sqrt{-g} R + A_{m},
\label{action}
\ee
where $R$ is the curvature scalar of four dimensional space-time
and and ${A_m}$ is the matter action, written in arbitrary 
curved coordinates. 
Gravitational field is identified with the metric tensor, 
${{ g_{\mu\nu},}}$ of the curved space-time. 
The curvature is created by matter through equations
of motion:
\be
R_{\mu\nu} -\frac{1}{2} g_{\mu\nu} R = 8\pi G_N T_{\mu\nu},
\label{GR-eqs}
\ee
where $R_{\mu\nu}$ is the Ricci tensor. There is no space here to 
stop on technicalities of Riemann geometry. A good introduction can be
found e.g. in book~\cite{teor-pol} where one can find definition and
properties of the Christoffel symbols, $\Gamma^\alpha_{\mu\nu}$,
Riemann tensor, $R_{\mu\alpha\nu\beta}$, Ricci tensor, 
$ R_{\mu\nu} = g^{\alpha\beta} R_{\mu\alpha\nu\beta}$, scalar curvature,
$R= g^{\mu\nu} R_{\mu\nu}$,
covariant derivatives in curved space-time, $D_\mu$, etc.

The source of gravity is the energy-momentum tensor of matter
taken in this curved space-time:
\be 
T_{\mu\nu} = 2 \delta A_{m}/\delta g^{\mu\nu}.
\label{T-mu-nu}
\ee
The impact of gravity on matter is included into $T_{\mu\nu}$
due to its dependence on metric and in some more complicated cases 
on the curvature tensors. Let us repeat that
the motion of matter in the gravitational field 
is simply the free fall, i.e. motion along geodesics.

Classical tensor theory of gravity agrees with all available data and
is a self-consistent, very beautiful and economic theory.
It is essentially based on one principle of general covariance, which
is a generalisation of Galilei principle of relativity to arbitrary
coordinate frames. 
Invariance with respect to general coordinate transformation 
(which is called general covariance) is a natural framework which
ensures vanishing of the graviton mass, $m_g$. 
Even if the underlying classical
theory is postulated to be massless, quantum corrections should generally
induce non-zero mass if they are not prevented from that by some
symmetry principle. This is another advantage of tensor gravity with
respect to scalar one for which no principle which forbids non-zero
mass is known. Though quantum gravity is not yet understood, it is
natural to expect that quantum corrections should induce $m_g \neq 0$
in absence of general covariance.

An important property of equations of motion (\ref{GR-eqs}) is that their
right hand side is covariantly conserved:
\be
D_\mu \left(R_{\mu\nu} -\frac{1}{2} g_{\mu\nu} R\right) \equiv 0.
\label{D-R-mu-nu}
\ee
Accordingly the energy-momentum tensor must be conserved too:
\be 
D_\mu T_{\nu}^{\mu\,(m)} = 0.
\label{DT}
\ee
Here, $D_\mu$ is covariant derivative, as we have already mentioned.
To those not familiar with Riemann geometry it may be instructive 
to mention that covariant derivative appears when one differentiates in
curved coordinate system, e.g. in spherical one, even in flat space-time.
From another point of view, covariant derivative in curved space-time,
which, e.g. is acting on vector field, looks as 
\be
D_\mu V_\nu  = \partial_\mu V_\nu - \Gamma^\alpha_{\mu\nu}  V_\alpha
\label{D-mu-V-nu}.
\ee
It is similar to covariant derivative in gauge theories,
because the latter includes gauge field, $A_\mu$ analogous to
$\Gamma_{\mu\nu}^\alpha$.

According to the Noether theorem,
the conservation of $T_{\mu\nu}$ follows from the least action principle
if the matter action is invariant with respect to general coordinate
transformation. So the gravitational (Hilbert) part of the action and
the matter part lead to self-consistent equations of motion only if
general covariance is maintained.

There is a deep analogy between the Einstein gravity and Maxwell
electrodynamics. The Maxwell equations have the form:
\be
\partial_\mu F^{\mu\nu} = 4\pi J^\nu
\label{max-eq}
\ee
Owing to anti-symmetry of ${ F^{\mu\nu}}$, the l.h.s. is automatically
conserved:
\be
\partial_\mu \partial_\nu F^{\mu\nu} \equiv 0,
\label{dF}
\ee
so the current must be conserved too:
\be
\partial_\mu J^\mu =0.
\label{dJ}
\ee 
These two conditions are consistent due to gauge invariance of the total
electromagnetic action with matter included.

Einstein was the first who decided to apply GR equations (\ref{GR-eqs})
to cosmology in 1918
and was very much disappointed to find that the equations 
do not have static solutions. So an advantage of GR was 
erroneously taken as a shortcoming. Only after the Friedman 
solution in 1922 which predicted the cosmological expansion~\cite{friedman} 
and the Hubble discovery of the latter in 1929~\cite{hubble}, the idea that 
our world is not stationary and may have a finite life-time was established.

{The distribution of matter in the
universe is assumed to be homogeneous and isotropic, 
at least in the early stage, as indicated by isotropy of 
cosmic microwave background radiation (CMB), 
and even now at large scales.}  Correspondingly the metric can be 
taken as {homogeneous and isotropic one (FRW metric~\cite{rw}):} 
\be
ds^2 = dt^2 - a^2(t)\,\left[f(r) dr^2 + r^2d\Omega \right],
\label{ds2}
\ee
where the function ${f(r)}$ describes 3D space of constant curvature,
${{ f(r) = 1/(1-k r^2)}}$. 

{The evolution of the scale factor ${{ a(t)}}$, i.e. the 
expansion law, is determined by the {Friedman equations,} 
which follow from the general GR ones for the FRW anzats.}
The derivation is straightforward but quite tedious.
We will derive them in the next section in very simple but not
rigourous way. The derivation may be taken as a mnemonic rule
to recall the equations in one-two minutes.

\section{Cosmological expansion \label{s-expan}}

\subsection{Basic cosmological equations \label{ss-eqs}}

Here we will present an oversimplified derivation of the Friedman equations.
Though the arguments are subject to criticism, the final results are
correct. Let us consider a test body on the surface of homogeneous 
sphere with radius $a(t)$ and the energy density ${\rho}$. 
The energy conservation condition for the non-relativistic test 
particle reads
${ v^2/2 = G_N M/a + const}$ where ${ M= 4\pi a^3 \rho/3} $. It can be
rewritten as:
\be
H^2 \equiv \left(\frac{\dot{a}}{a}\right)^2 =
 \frac{8\pi\,\rho\, G_N }{3} - \frac{k}{a^2}
\label{H2}
\ee
This is one of the main cosmological equations. Here is the famous
Hubble expansion law, that the object situated at distance $d$ runs away
from us with velocity proportional to the distance, $v= Hd$. Notice that
for $d>1/H$ it runs away with superluminous velocity. Superluminous velocities 
of distant objects are allowed by GR but locally velocities must be always 
smaller or equal to the speed of light. 

Another equation follows from the
energy balance of the medium inside the sphere: ${ dE = -P\,dV}$
where 
${ E =\rho V}$ and $ { dE = V d\rho + 3 (da/a) V\rho  }$.
Hence: 
\be
\dot{\rho}+3H(\rho+P)=0.
\label{dot-rho}
\ee
This equation is simply the law of covariant energy-momentum 
conservation (\ref{DT}) in metric (\ref{ds2}).

{\it Problem 1.} {Derive from eqs. (\ref{H2}) and (\ref{dot-rho}) 
the law for the acceleration of the test body:}
\be
\frac{\ddot a}{a} = -\frac{4\pi\,G_N }{3}\,(\rho+3P)
\label{ddot-a}
\ee

A striking feature of equation (\ref{ddot-a}) is that 
not only energy but also
pressure gravitates. It is always assumed in canonical theory that
$\rho$ is positive, though pressure may be negative. Thus if
$\rho + 3P <0$, the cosmological expansion would proceed with acceleration,
i.e. antigravity may operate in cosmological scales.
In other words, negative pressure is the source of the cosmological 
expansion. Life is possible only because of that. We believe that the
universe was in such anti-gravitating state at the very beginning, during the
so called inflationary stage (see below). Surprisingly it was established
during the last decade that at the present time the expansion is also
accelerating. There existed a simple analogy between the universe expansion
and the motion of a stone thrown up from the Earth with some initial velocity
$v_0$. The speed of the stone drops down with time and it either 
will return back
to the Earth, if $v_0$ is smaller than a certain value, $v_1$. If $v_0>v_1$
the stone will never come back. In the last case the stone will either come
to infinity with non-zero speed or with the vanishing one. All three such regimes
could exist in cosmology and they were believed to be realised, depending
upon the initial expansion velocity. The first regime corresponds to the 
closed universe with $\rho> \rho_c$, where $\rho_c$ is the critical or closure
energy density, see below.
The expansion in this case will ultimately turn into contraction.
The other two regimes correspond to open universe which was expected to 
expand forever. The third one with zero velocity at infinity corresponds
to spatially flat universe, with $k=0$ in eq. (\ref{H2}). Now the picture 
is very much different. Imagine that you have thrown a stone from the Earth
and first the stone moves with normal negative acceleration and after
a while starts to move faster and faster as if it has a rocket engine. 
This is exactly what we see in the sky now. It means, in particular, that
the spatially closed universe may expand forever. 
This accelerated cosmological expansion, induced by antigravity at cosmological
scale is prescribed to existence of mysterious dark energy. It is one the greatest
unsolved problems in  modern fundamental physics.

{\it Problem 2.} We seemingly started from the Newtonian theory
but came to the conclusion that pressure gravitates which is not
the true in Newtonian case. Where is the deviation from Newton?\\ 
{\it Problem 3.} Prove that for positive definite energy density, $\rho > 0$,
any object of finite size creates an attractive gravitational force,
even if inside such an object pressure may be arbitrary negative.\\
{\it Problem 4.} Prove that any finite object with positive energy density
gravitates, so antigravitational action of pressure can manifest itself only in
infinitely large objects.

Except for equations which determine the law of the cosmological 
expansion, we need an equation which governs particle propagation
in FRW metric, i.e. geodesic equation. The latter can be written as:
\be 
\frac{d V^\alpha}{ds} = - \Gamma_{\mu\nu}^\alpha V^\mu V^\nu + 
{\rm curvature \,\,\, term},
\label{dV-ds}
\ee
where ${V^\alpha = dx^\alpha/ds}$ and the curvature term
is absent in spatially flat universe, when $k=0$. In what follows
we will consider only this case, moreover, the effects of curvature
are typically small.

To solve this equation one needs first to calculate the Christoffel symbols
for metric (\ref{ds2}). In 3D flat space they have very simple form:
\be
\Gamma^{i}_{jt} = H \delta^i_j,\,\,\, 
\Gamma^{t}_{ij} = H a^2\delta_{ij}.
\label{Gamma}
\ee
All other are zero. After that the geodesic equation takes 
a very simple form:
\be
\dot p = -Hp
\label{dot-p}
\ee
with an evident solution $p \sim 1/a(t)$ which describes red-shifting of
momentum of a free particle moving in FRW background. In derivation of this
result one has to pay attention that physical momentum is defined with
respect to physical length $dl = a(t) dx $.

We can simply derived the same result taking into account the
Doppler red-shift of the momentum of a free particle induced by
the cosmological expansion.
Let us take two points A and B separated by distance $dl$.
The relative velocity of these two points due to expansion is
{${ U= H dl}$}. The Doppler shift of the momentum
of the particle moving from A to B with velocity $v=dl/dt$ is
\be
dp = -UE = -HE dl.  
\label{dp}
\ee
Thus
\be 
\dot p = -HE dl/dt = -Hp.
\label{dot-p-Dopp}
\ee
Let introduce at this stage the notion of the cosmological
red-shift:
\be
z = a(t_U)/a(t) - 1,
\label{z}
\ee
where $t_u$ is the universe age and
$a(t_U)$ is the value of the scale factor today. So the
momentum of a free particles drops down in the course of expansion as
{${ p \sim 1/a \sim 1/(z+1)}$.}

Equations (\ref{H2}-\ref{ddot-a}) are the basic cosmological
equations for three unknowns, $a$, $\rho$, and $P$. 
However, there are only two independent equations. So it is necessary
to have one more equation describing properties of matter i.e.
the equation of state (e.o.s.): 
$P=P(\rho)$. Usually this equation is parametrized 
in the simple linear form:
\be 
P = w \rho.
\label{P-of-rho}
\ee
The parameter $w$ determines matter properties. For non-relativistic matter
pressure is negligibly small in comparison with $\rho$ and in a good
approximation we can take $w=0$. For relativistic matter $w=1/3$.
There is also one more type of matter (or vacuum) known to exist in
the universe, for which $w=-1$. 

However, sometimes the equation of state does not exist but the necessary 
additional relation (not e.o.s.)
can be derived from the equations of motion. 
E.g. for a scalar field:
\be 
D^2 \phi + U'(\phi) = 0
\label{D2-phi}
\ee
and one can calculate ${ T_{\mu\nu}}$ and find ${ \rho}$ and ${ P}$
but ${ P\neq P(\rho)}$.

{\it Problem 5.} 
Calculate ${ T_{\mu\nu}(\phi)}$, ${ \rho}$, and ${ P}$ for homogeneous field
$\phi(t)$.

\subsection{Expansion regimes \label{ss-expan}}

Here we will present solutions of the cosmological equations
for several special cases which were/are realised in the universe
at different stages of her evolution. We always assume that the 
three dimensional space is flat, i.e. $k = 0$. As we see below it is true
during practically all life-time of the universe.

Let us first consider non-relativistic matter with equation of state $P=0$. 
According to eq. (\ref{dot-rho}) the evolution of the energy density is 
given by:
\be
\dot \rho = -3H\rho
\label{dot-rho-nr}
\ee
and thus ${ \rho\sim 1/a^3}$. The result is evident, it is simply
dilution of the number density of massive particle at rest.

The time dependence of the cosmological scale factor, is determined by
eq. (\ref{H2}):
\be 
\dot a/a \sim \sqrt{\rho}
\label{dota-nr}
\ee
and thus in non-relativistic regime {${ a\sim t^{2/3}}$} and $H=2/3t$.

For relativistic matter the equation of state is {${ P=\rho/3}$}
and correspondingly
\be
\dot \rho = -4H\rho .
\label{dot-rho-rel}
\ee
Thus $\rho$ drops as ${ \rho\sim 1/a^4}$ and the scale factor rises as
$a(t) \sim  t^{1/2}$, which means that $H=1/2t$.

The energy density of relativistic 
particles drops one power of $a$ faster than that of non-relativistic ones
due to dilution of their number density as volume, $1/a^3$, and red-shift of
the particle momentum. That's why relativistic matter dominated in the
early universe, while at  a later stage non-relativistic matter took over.
Until last years of the XX century it was believed that the universe today
is dominated by non-relativistic matter but then it was established that
the dominant matter is the so called dark energy with equation of state
close to the vacuum one.

In the vacuum(-like) regime the energy-momentum tensor is proportional
to the  metric  tensor which is the only invariant tensor:  
\be
T_{\mu\nu} =\rho_{vac}\,\, g_{\mu\nu}.
\label{T-vac}
\ee
Hence $P_{vac} = -\rho_{vac}$ and vacuum energy density remains constant 
in the course of the cosmological expansion:
{${ \dot \rho = -3H(\rho+P) = 0}$}. The scale factor in this case rises 
exponentially ${ a\sim \exp (Ht)}$.

According to our understanding, all visible universe originated 
from microscopically small volume with
negligible amount of matter by exponential expansion with practically constant
${\rho}$, see below.

It is interesting to calculate the causality distance as a function of time,
which is equal to the light path propagating from an initial moment $t_1=0$ to final
moment $t_2$. This distance can be found from the light geodesic equation:
$dt^2 - a^2(t) dr^2 =0$. It can be easily integrated to give:
\be
l_\gamma = a(t)\,\int_0^t \frac{dt'}{a(t')}\,.
\label{l-gamma}
\ee
Thus for relativistic regime, $l_\gamma = 2t$, for non-relativistic one it is
${l_\gamma = 3t}$, and for exponential De Sitter (inflationary stage):
$l_\gamma = H^{-1} [\exp(Ht) - 1]$.

As we see, cosmological equations do not have
stationary solutions, at least for the examples taken.
In the case of positive space curvature, i.e. ${ k>0}$, and 
normal matter  with ${ \rho \sim 1/a^n}$, n=3,4, the expansion will 
ultimately change into contraction. If however,
${\rho >k/a^2}$, e.g. if $\rho$ is vacuum energy, 
the expansion  may last forever for any $ k$.

Note, that if the cosmological energy density is dominated by the normal
matter the Hubble parameter drops down as $H\sim 1/t$, where $t$ is the
universe age. If the universe is dominated by vacuum(-like) energy, the
Hubble parameter remains constant, 

{\it Problem 6.} Find ${ \rho (a)}$ and ${ a(t)}$ for general linear 
equation of state, $P=w\rho$ with arbitrary $ w$. Study the case of 
${ w \leq -1}$.

\subsection{Cosmological parameters \label{ss-cosm-par}}

Before proceeding further let us say a few words about the natural system
of units which is used throughout all these lectures. We take speed of light,
reduced Planck constant, and Boltzmann constant all equal to unity. 
{${ c=h/2\pi = k =1}$}. All dimensional quantities have dimension of 
length, or time, or (inverse) mass or energy -- all the same. For example
the Newtonian gravitational constant has dimension of inverse mass, 
{${ G_N \equiv 1/M_{Pl}^2};$} ${ M_{Pl} = 1.221 \cdot 10^{19}\,{\rm GeV}=
2.176\cdot 10^{-5}\,{ g}}$;
{${ m_p = 938\,{\rm MeV} = 1.67\cdot 10^{-24}\, {\rm g}}$;}\\
${1\,{\rm GeV}^{-1}} = 1.97\cdot 10^{-14}\,\,{\rm cm} 
= 0.66 \cdot 10^{-24}, {\rm{s}}$;
{${1\, {\rm eV} = 1.16\cdot 10^4 K^o}$.}

Homogeneous cosmology is described in terms of the Hubble parameter
{${ H = \dot a/a}$,} which characterises the universe expansion rate,
by the critical or closure energy density 
\be
{ \rho_c = 3H^2\, M^2_{Pl}/8\pi}
\label{rho-c}
\ee
and by the dimensionless parameter {${ \Omega_j = \rho_j/\rho_c}$,}
which measures the relative contribution of the energy density of species of 
type $j$ into the total energy density of the universe.
Clearly for spatially flat universe
the total energy density is equal to the critical one:
{${ \Omega_{tot} = 1}$, if ${ k=0}$}. It remains constant in the course of the
universe expansion.

If $k\neq 0$, then from eq. (\ref{H2}) follows that 
${ \Omega}$ evolves with time as
\be
\Omega (a) = \left[ 1- \left( 1 - \frac{1}{\Omega_0}\right)\,
\frac{\rho_0 a_0^2}{\rho a^2}\right]^{-1},
\label{Omega-of-t}
\ee
where the index sub-0 denotes the present day values of the corresponding 
quantities.

For normal matter ${ \rho a^2 \rar 0}$ if  ${ a\rar \infty}$ and
${\Omega}$ runs away from unity:
{${ \Omega(a)\rar 0}$ if ${ \Omega_0 < 1}$} and
{${ \Omega(a)\rar \infty}$ 
if ${ \Omega_0 > 1}$.} On the other hand,
{${ \Omega(a)\rar 1}$ when ${ \rho a^2 \rar \infty},$}
e.g. for vacuum energy at expansion {for arbitrary initial 
${\Omega}$.}

The present day value of the $H$ characterises by dimensionless parameter
$h$ as
\be
H= 100\, h\,{\rm km/sec/Mps},
\label{H-0}
\ee
where ${ h= 0.73 \pm 0.05}$. The inverse quantity
${H^{-1}= 9.8\,{\rm Gyr}/h \approx 13.4\, {\rm Gyr}}$ is 
approximately equal to the universe age.

An exact expression for the universe age through the present day values of 
the Hubble parameter and relative energy densities of different forms of matter
can be obtained by integration of the equation
\be
\dot a  =\left[
 {8\pi\,\rho\, G_N\, a^2}/{3} - {k}\right]^{1/2}
\label{dot-a2}
\ee
After simple algebra one finds: 
\be
t_U = \frac{1}{H}\,\int_0^1 \frac{dx}
{\sqrt{1-\Omega_t +\frac{\Omega_m}{x}
+\frac{\Omega_r}{x^2} + x^2\Omega_v }},
\label{t-U}
\ee
where $\Omega_t$ is the total $\Omega$ and $\Omega_{r,m,v}$ are respectively
contributions from relativistic and non-relativistic matter and from
vacuum energy. All the quantities in this equations are the present day ones; we
skipped the sub-index 0 here.

{\it Problem 7.}  Derive eq. (\ref{t-U}). 
Find ${ t_U}$ for ${\Omega_t = \Omega_m = 0,\,\,0.3,\,\,1}$.
Find ${ t_U}$  for ${\Omega_t = 1}$, ${ \Omega_m = 0.3}$, and 
${ \Omega_v = 0.7}$.

The ``measured'' value of the universe age lies in the interval
\be
t_U = 12-15 \,{\rm Gyr},
\label{t-U-mes}
\ee
found from the ages of old stellar clusters and nuclear chronology.

{The present day value of the critical energy density is :}
\be
\rho_c = \frac{3H^2 m_{Pl}^2}{8\pi} =
1.88\cdot 10^{-29} h^2 {\rm { \frac{g}{cm^3}}}=
10.5\, h^2  {\rm { \frac{\rm keV}{\rm cm^3}}} 
\approx 10^{-47} h^2\,{\rm { GeV^4}}
\label{rho-c-0}
\ee 
It corresponds approximately
to 10 protons per ${m^3}$, but the dominant matter
is not the baryonic one and in reality there are about 0.5 protons per 
cubic meter.

\subsection{Matter inventory \label{ss-inventory}}

The relative contributions of different forms of matter
into the total energy density are obtained from different independent
astronomical observations. Here we only present their numerical values.
For discussion of their measurements in more detail a few extra 
lectures are necessary.

The total cosmological energy density is very close to the critical one:
${{ \Omega_{tot} = 1 \pm 0.02}}$ as found 
from the position of the first peak of  the angular spectrum of CMBR and 
the large scale structure (LSS) of the universe.

The usual baryonic matter makes quite small contribution: 
${{ \Omega_{B} = 0.044 \pm 0.004}}$ as found 
from the heights of the peaks in angular fluctuations of CMB, 
from produciton of light elements at BBN, and from the onset of
structure formation with small ${ \delta T/T}$. 

Approximately five time more than baryons  is brought by the so called
dark matter. It is invisible matter with presumably normal gravitational 
interactions. As is found 
from galactic rotation curves, gravitational lensing, equilibrium
of hot gas in rich galactic clusters, cluster evolution, and LSS:
{${{ \Omega_{DM} \approx 0.22\pm 0.04 }}$}.

The rest, ${{ \Omega_{DE} \approx 0.76,}}$, is carried by some mysterious 
substance, which is uniformly distributed in the universe
induces accelerated cosmological expansion. Its equation of state is
close to the vacuum one, i.e. ${{ w \approx -1}}$. The existence and the
properties of dark energy was deduced from 
dimming of high-z supernovae, LSS,  CMB spectrum, and the universe age.

I would like to stress that
the different pieces of data and their interpretation are
{independent.} It minimises the probability of a
possible interpretation error. The numerical values obtained in different
type measurements are pretty close to each other.

\section{Brief cosmological history \label{s-history}}

Universe history can be separated into several epochs, some of them are
described by established well known physics verified by experiment, 
some are based on hypothetical physics beyond the standard model, and some
(little?) are absolutely dark. 
\begin{enumerate}
\item{}Beginning, unknown. 
Quantum gravity, quantum space-time?
Maybe time did not exist? It is so called pre-inflationary cosmology.
\item{}Inflation, i.e. epoch of exponential expansion of the universe. 
It is practically ``experimental'' fact..
\item{} End of inflation, particle production. At that period dark expanding
``emptiness'' filled by scalar (or some other) field, inflaton, exploded,
creating light and other elementary particles.
\item{} Baryogenesis. At that time an excess of matter and antimatter in
the universe (or vice versa) was created.
\item{} Thermally equilibrium universe, adiabatically cooled down. Presumably during
this epoch several phase transitions took place leading to breaking of 
grand unified symmetry (GUT), electroweak (EW) symmetry, supersymmetry
(SUSY), phase transition from free quark-gluon phase to confinement phase in   
quantum chromodynamics (QCD), {\it etc.} with possible formation
of topological defects and non-topological solitons. At the phase transitions
adiabaticity of expansion could be broken.
\item{} Decoupling of neutrinos from electromagnetic part of the cosmological 
plasma. It took place when the universe was about 1 second old 
at ${ T\sim 1}$ MeV.
\item{} Big bang nucleosynthesis (BBN), which proceeded in the time interval from
1 s to $\sim$ 200 s, and ${ T=1-0.07}$ MeV. At that time light elements, $^2H$,
$^3He$, $^4He$, and $^7Li$ were formed. Theory is in a good agreement with 
observations. A different mechanisms for creation of light elements is unknown.
It makes BBN  one of the cornerstones of the standard cosmological model
(SCM). 
\item{} Onset of structure formation which started when the dominating cosmological
matter turned from relativistic into non-relativistic one. 
It took place at the red-shift ${ z_{eq} \approx 10^4}$, ${ T \sim }$ eV.
\item{} Hydrogen recombination, at ${ z\approx 10^3}$ or
$ T \sim 0.2$ eV. At that time cosmic microwave radiation (CMB)
decoupled from matter and after that it propagated practically freely in 
the universe. After decoupling of matter and radiation
baryons begun to fall into already evolved seeds of structures
created by dark matter (DM).
\item{} Formation of first stars and reionization of the universe.
\item{}Present time, ${ t_U = 12-15}$ Gyr.
\end{enumerate}

\section{Hot equilibrium epoch \label{s-hot}}

Usually a system comes to the state of thermal equilibrium after 
sufficiently long time. Paradoxically, in cosmology equilibrium is reached 
in the early universe when time is short but temperature is high and the
reaction rates $\Gamma$
exceed the cosmological expansion rate, $H=\dot a /a$: 
\be
\Gamma = \sigma n \sim \alpha^2 T > H \sim T^2/M_{Pl}
\label{Gamma-H}
\ee
This condition is fulfilled at high temperatures but bounded by
${ T \leq \alpha^n M_{Pl}}$, where $\alpha$ is the generic value of the
coupling constant and $n=1,2$ for decays and reactions respectively.
At lower $T$ the equilibrium is broken due to Boltzmann suppression of the
participating particles.

In equilibrium particle distribution functions are determined by two
parameters only, by the temperature, $T$ and, if the particles do not
coincide with antiparticles, by their chemical potential, $\mu$.
{The equilibrium distributions have the well known form:}
\be
f^{(eq)}_{f,b}(p) =\frac{1}{ \exp \left[ (E-\mu )/T \right] \pm  1 },
\label{f-eq}
\ee
where ${ E=\sqrt {p^2 +m^2}} $. 
Equilibrium  with respect to the reaction
${ a_1 +a_2  +a_3  \ldots \leftrightarrow b_1 +b_2 +\ldots} $
imposes the following condition on the chemical potential of the 
participating particles: 
\be 
\sum_i \mu_{a_i} =\sum_j \mu_{b_j}
\label{summu}
\ee
Chemical potentials are necessary to introduce in
charge asymmetric case to describe inequality between number densities of
particles and antiparticles:  ${n\neq \bar n}$. It is assumed usually 
that in cosmological situation chemical potentials are very small, as follows
from the observed baryon asymmetry of the universe. However, large lepton
asymmetry is not excluded and so it may be interesting to consider
non-negligible chemical potentials. One should keep in mind however, that
chemical potential of bosons cannot be arbitrarily large. It is
bounded by the mass of the boson, $\mu < m$, while for fermions there is
no upper limit on $\mu$.

What happens if charge asymmetry in bosonic sector, i.e.
{${ (n -\bar n)}$} is so large that  {${ \mu = m}$} is not 
sufficient to realise that? In this case the equilibrium distribution 
function acquires an additional term:
\be
f = \left[e^{(E-m)/T} -1\right]^{-1} + C\delta^3 (p)\,, 
\label{bose-cond}
\ee
i.e. Bose condensate forms. Notice that equilibrium distributions are
always determined by two parameters: 
${ T}$ and ${ \mu }$ if $\mu<m$ or 
${ T}$ and ${C }$ if $\mu$ is fixed by the maximally allowed value,
$\mu = m$.

{\it Problem 8.} Check that the distribution (\ref{bose-cond}) is indeed an 
equilibrium solution of kinetic equation, i.e. ${ I^{coll} =0}$. 

If the reactions of annihilation of particle and antiparticle
\be
b +\bar b \lrar 2\gamma\,, 3\gamma
\label{annih}
\ee
are in equilibrium, then from condition (\ref{summu})
follows that $\mu_\gamma = 0$ and that the chemical potentials of
particles and antiparticles are equal by magnitude and have opposite
signs:
\be
\mu + \bar \mu =0.
\label{mumubar}
\ee
If the equilibrium with respect to annihilation into two photons 
is maintained, while the annihilation into larger number of photons
is out of equilibrium (such reactions are slower due to an extra 
power of the fine structure constant $\alpha$ and smaller phase space), 
non-zero chemical potential of photons can be developed. The observed
CMB photons have zero or very small chemical potential $|\mu/T| < 10^{-4}$.

{The equilibrium number density of bosons with ${ \mu =0}$} is:
\be
n_b 
\equiv \sum_s \int \frac{f_b (p) }{(2\pi)^3}  \,d^3p  =
\left\{ \begin{array}{ll}{
\zeta  (3)  g_sT^3/  \pi^2  \approx  0.12g_sT^3,\,\, 
 T>m;} \\
 (2\pi)^{-3/2}g_s (mT)^{3/2} e^{-m/T},\,\, T<m\,,
\end{array}\right.
\label{nb}
\ee
where ${ g_s}$ is the number of spin states.

The number density of photons is equal to:
\be 
n_\gamma =0.2404T^3=412 (T/2.728 {\rm K})^3\,{\rm cm}^{-3}\,,
\label{n-gamma}
\ee
where {2.728 K is the present day temperature of the cosmic microwave
background radiation (CMBR).}

The equilibrium number density of non-degenerate (i.e. $\mu =0$) fermions is: 
\be{
n_f=}\left\{  \begin{array}{ll}
{ 3 n_b/4  \approx  0.09\,g_s\,T^3,\,\,  T>m;}  \\
{ (2\pi)^{-3/2}g_s (mT)^{3/2} e^{-m/T},\,\, T<m.
}\end{array}\right.
\label{nf}
\ee

{The equilibrium energy density is given by:}
\be
\rho =\sum \frac{1}{2\pi^2} \int \frac{\,dpp^2E}{ \exp[(E-\mu)/T] \pm  1}.
\label{rhoeq}
\ee
The total energy density of all species of relativistic matter with $\mu =0$ is
\be
\rho_{rel} =(\pi^2/30)g_*T^4\,,
\label{rhorel}
\ee
where ${ g_*= \sum \left[g_b +(7/8)g_f\right]}$ and $g_{b,f}$ is the 
number of spin states of bosons or fermions.
{\it Problem 9.} 
Calculate ${g_* }$ for ${T \sim 3}$ MeV. {Answer:  10.75.} 

{Sometimes the total energy density is described by expression (\ref{rhorel})
with the temperature depending number of species, 
${ g_* (T)}$ which includes contributions of all relativistic as
well as non-relativistic species.}

The energy density of CMB photons is
\be
\rho_\gamma  =\frac{\pi^2 T^4}{  15}   \approx   
0.2615\left( \frac{T} {2.728\, {\rm K}}\right)^4 \frac{\rm eV } { cm^3} 
 \approx 4.662\cdot 10^{-34}
\left( \frac{T} {2.728{\rm{K}}}\right)^4  {\frac{{\rm g} }{ {\rm cm}^3}}\,.
\label{rhogamma}
\ee
The relative contribution of CMB into the cosmological energy density
is small but non-negligible:
\be
\Omega_{CMB} = 4.7\cdot 10^{-5}\,.
\label{Omega-CMB}
\ee

Heavy particles, i.e. those with ${ m>T}$, have exponentially small
number and energy densities if they are in equilibrium:
\be
\rho_{nr} =g_s m\left( \frac{mT}{ 2\pi}\right)^{3/2}
e^{-m/T}
\left(1+
\frac{27T}{ 8m} +\ldots \right)
\label{rhonr}
\ee
{If the annihilation stopped, the density of massive particles
could strongly exceed the equilibrium one.} If the particles are 
unstable with a large life-time then at later stage their distribution 
would return to the equilibrium one.

Approach to equilibrium and deviations from it in homogeneous 
cosmology are described by the kinetic equation in FRW space-time:
\be
\frac{df_i}{dt} = (\partial_t + \dot p \partial_{p_i}) f_i =
(\partial_t - H\,p_i \partial_{p_i}) f_i = I_i^{coll}\,,
\label{kin-eq}
\ee
where ${ \dot p = - Hp}$ and
$I_i^{(coll)}$ is the collision integral, see below eq. (\ref{I-coll}).

{\it Problem 10.} Why in the distribution function $ p$ and $ t$ are taken
as independent variables, while above we treated momentum as a function of time,
${p = p(t)}$? 

In terms of dimensionless variables:
\be
x= m_0 a\,\,\, {\rm and}\,\, y_j = p_j a
\label{xyi}
\ee
the l.h.s. of kinetic equation takes a very simple form:
\be
Hx\frac{\partial f_i}{\partial x } = I^{coll}_i\,.
\label{hxdfdx}
\ee

If the universe is dominated by relativistic matter, the 
temperature drops  as ${T\sim 1/a}$ and the Hubble parameter is 
expressed through $x$ as:
\be
H = 5.44 \sqrt{{g_* \over 10.75}}\, {m_0^2 \over x^2 m_{Pl}}.
\label{H-of-x}
\ee
In thermal equilibrium:  ${ g_*=2}$ for photons, 
${g_*=7/2}$ for ${ e^{\pm} }$-pairs, and ${ g_*=7/8}$
for one family of left-handed neutrino.
Since $H=1/2t$ the relation between cosmological time and temperature of the
primeval plasma has the form:
{${ t/{\rm sec} \approx ({\rm MeV}/T)^2}$.}

For non-interacting particles, i.e. for ${ I^{coll} = 0}$, 
equation
\be
Hx{\partial f \over \partial x } = 0
\label{Hxdxf}
\ee
is solved as
\be
f= f(y, x_{in})\,,
\label{f-nonint}
\ee
where $x_{in}$ is an initial value of the scale factor. Thus the distribution
function maintains its initial form in terms of variables $x$ and $y$.
For massless particles with non-zero chemical potential:
{${ f = f_{eq} (y,\xi)}$}, if they ever were in 
equilibrium. Here {${ T\sim 1/a}$} and $ { \xi = \mu/T = const}$.
{Initially equilibrium distribution of massless particles 
maintains its equilibrium form even after 
interactions are switched off,} as is observed in {CMB}.

Let us check this important statement. The l.h.s. of the kinetic 
equation can be written as: 
\be
 \left(\partial_t - Hp \partial_p \right) 
f_{eq}\left[\frac{E-\mu(t)}{T(t)}\right] = 
\left[-\frac{\dot T}{T}\,\frac{E-\mu}{T} -\frac{\dot \mu}{T}  
-\frac{Hp}{T}\right]  \frac{df_{eq}}{dx} .
\label{lhsm-zero}
\ee
The factor in square brackets vanishes if 
$\dot \mu = \dot T/T= -H$, which is true in the expanding universe, 
and if $ E(\dot T /T) = -Hp$ which can be and is true only for  $E=p$
i.e. for $ m=0$.

If the particle mass is non-zero and if 
the interaction is switched off at ${ T\gg m}$, the distribution
looks as an equilibrium one but in terms of {{${ p/T}$} but not
{${ E/T}$.}

{\it Problem 11.} Find the distribution of massive particles decoupled at
${ T<m}$. What if decoupling is non-instantaneous?

In the equilibrium state and for vanishing chemical potentials
entropy, $S$, in comoving volume is conserved:
\be
\frac{dS}{dt} \equiv
\frac{d}{dt}\,\left(a^3\,\frac{\rho+P}{T}\right) = 0
\label{dS-dt}
\ee
In fact this equality is more general. It is
true for any distribution function {${ f = f(E/T)},$}
satisfying the condition of the covariant energy conservation,
${ \dot \rho = -3H \left(\rho + P\right)}$ with arbitrary $T(t)$.
So we find:
{\be
{{ \frac{d}{dt}\,\left(a^3\,\frac{\rho+P}{T}\right)= }} 
{{ a^3\,\left[\frac{\rho+P}{T}\,\left(3H - \frac{\dot T}{T} - 3H\right)
+ \frac{\dot P}{T}\right],}
}\label{ds-dt2}
\ee}
where the pressure is:
\be
P = \int \frac{d^3 q}{(2\pi)^3}\, \frac{q^2}{3E}\, f\left(\frac{E}{T}\right)\,.
\label{P}
\ee
From this expression we can find ${ \dot P}$ (remember that only ${ T}$ 
depends upon time) and integrate by parts to obtain:
\be
\dot P = \frac{\dot T}{T}\,(\rho + P)\,,
\label{dot-P}
\ee
which leads to the conservation law (\ref{dS-dt}).

Let us return now to the kinetic equation and specify the 
collision integral for an arbitrary process: 
${ i+Y \lrar Z}$:
\be 
I^{coll}_i=\frac{(2\pi)^4}{2E_i}  \sum_{Z,Y}  \int  \,d\nu_Z  \,d\nu_Y
\delta^4 (  p_i +p_Y -p_Z)
\lbrack |A(Z\rightarrow  i+Y)|^2
\prod_Z f \prod_{i+Y} (1\pm f) - \\ \nonumber 
|A(i+Y\rightarrow  Z)|^2
 f_i  \prod_Y  f\prod_Z  (1\pm  f)\rbrack \,,
\label{I-coll}
\ee
where 
${Y}$ and ${Z}$ are  arbitrary, generally  multi-particle  states,
$\prod_Y f$ is  the  product  of  phase space densities  of  particles
forming the state $Y$, and
\be
d\nu_Y = 
\prod_Y {d^3p\over (2\pi )^3 2E}
\label{dnuy}
\ee
The signs '+' or '$-$' in ${ \prod (1\pm f)}$ are chosen for  bosons  and
fermions respectively.

{Equilibrium distributions} by definition annihilate the 
collision integral:
\be
I^{coll} [f^{(eq)}] = 0
\label{I-coll-0}
\ee
The standard Bose/Fermi distributions do that.
It is easy to check that this is indeed true in
T-invariant theory where the detailed balance condition holds,
\be
{{|A_{if}(p)|^2 = |A_{fi}(p')|^2}}
\label{Aif-Afi}
\ee
where $p'$ is time reversed momentum, i.e. with the opposite sign of the
space coordinate with respect to $p$.

So after an evident change of variables in the collision integral
we can factor out $|A_{if}|^2$ and the integrand would be
proportional to:
\be
\Pi f_{in} \Pi (1\pm f_{fin}) - \Pi f_{fin} \Pi (1\pm f_{in})=0
\label{f-in-f-fin}
\ee
It is easy to check that the functions ${ f_{eq}}$ (\ref{f-eq})
annihilate the collision 
integrals due to {conservation of energy}
\be
\sum E_{in} = \sum E_{fin}
\label{E-consrv}
\ee
and if chemical potentials satisfy:
\be
\sum \mu_{in} = \sum \mu_{fin}.
\label{mu-consrv}
\ee
This condition is {enforced} by reactions.

Since we know that CP-invariance is broken and (mostly) believe that CPT
invariance holds, we must conclude that the invariance with respect to 
time reversal, T-transformation, is broken as well. It means, in particular,
that the detailed balance condition is invalid, 
$|A_{if}|^2 \neq |A_{fi}|^2$. Now a natural question arises:
would the usual equilibrium distributions survive  
in T-violating theory? Let us check what happens with the collision integral
for $f=f_{eq}$. Due to eq. (\ref{f-in-f-fin}) the integrand is proportional to
\be
I_{coll} \sim \Pi f_{in}(1\pm f_{fin})
\left( |A_{if}|^2 - |A_{fi}|^2 \right)\,.  
\label{I-coll-T-noninv}
\ee
The last factor is {non-vanishing} if $T$-invariance is broken.
However, due to S-matrix unitarity (or hermicity of the Hamiltonian)
breaking of $T$-invariance is observable only if several processes 
participate and though each separate term is non-zero, the sum over 
all relevant processes vanishes~\cite{ad-cyclic}.

It can be proven using S-natrix unitarity condition, 
$ S S^\dagger =1$. If as usually we introduce scattering matrix:
$ S= (I +iT)$, then it satisfies:
\be
i(T_{if}-T_{fi}^\dagger ) = 
- \sum_n\, T_{in} T^\dagger_{nf}
= -\sum_n\, T^\dagger_{in} T_{nf}
\label{unit-T}
\ee
Summation over $n$ includes integration over phase space.
Instead of detailed balance a new condition of cyclic 
balance~\cite{ad-cyclic} 
\be
\sum_k\int d\tau_k \left(|A_{ki}|^2 - |A_{ik}|^2\right) =0
\label{cycl}
\ee
ensures vanishing of ${ I_{coll}}$ on ${ f=f_{eq}}$.
Here ${d\tau_k}$ includes Bose/Fermi
enhancement/suppression factors.
For validity of this relation full unitarity is not necessary.
Normalisation of probability $\sum_f w_{if} =1$
plus CPT invariance are sufficient.
If CPT is broken, then the additional condition 
$\sum_f w_{if} =1,\,\,\, \sum_f w_{fi} =1$ would save the standard 
equilibrium statistics. However, in the case that 
nothing above is true, equilibrium distributions 
would differ from the canonical ones. If the theory does not
respects sacred principles of unitarity, hermicity, etc., the Pandora box 
of disasters would be open and equilibrium might deviate very much
from the standard case or even not exist.

\section{Freezing of species \label{s-freeze}}

As we discussed in the previous section, primordial plasma is typically in 
thermal equilibrium state in the early universe.
When $T$ dropped down to the so called decoupling temperature, ${T_d}$,
(sometimes it is called freezing temperature, ${T_f}$) the 
interaction with plasma effectively switched off and the 
particles started to behave as free, non-interacting ones.
{There are two types of decoupling:}\\
{1. Relativistic freezing, when ${T_d > m}$. This is realised e.g. for
neutrinos, or some other hypothetical 
weeakly interacting particles. Such particles by definition
make hot dark matter (HDM) or warm dark matter (WDM).}\\
{2. Non-relativistic freezing, when ${T_d < m}$. Such particle make 
cold dark matter (CDM).}

\subsection{ Relativistic freezing \label{ss-rel-freeze}}

{Neutrinos decoupled at temperatures much larger than their mass,
${ T \gg m_\nu}$.} It can be seen from the decoupling condition that the weak
interaction rate became smaller than the expansion rate ${ \sigma_W\, n < H}$ at:
\be 
G_F^2 E^2\,T^3 \sim T^2/M_{Pl}\,,
\label{weak-decpl}
\ee
where $G_F = 1.166\times 10^{-5}$ GeV$^{-2}$ is the Fermi coupling constant.

Since the energy of relativistic particles is equal by an order of magnitude
to the plasma temperature ${E\sim T}$ we obtain:
\be 
T_f \sim m_N\, \left(10^{10}m_N/ M_{Pl}\right)^{1/3} \sim {\rm MeV}.
\label{T-f-nu}
\ee
More accurate calculations are necessary to establish if 
${T_f}$ is larger or smaller than ${m_e}$, which is {important for
the calculations of the number density, ${n_\nu}$, of relic neutrinos
at the present time and for the cosmological bound on their mass, ${m_\nu}$.}

For more accurate calculations of the decoupling temperature from the 
electron-positron plasma we will use the 
kinetic equation in Boltzmann approximation with only direct reactions
with electrons, i.e.
${ \nu e}$ elastic scattering and ${ \nu \bar\nu}$-annihilation taken into 
account:
\be
Hx\,{\partial f_\nu \over f_\nu\,\partial x} =
-{80G_F^2\left( g_L^2 + g_R^2 \right) y \over 3\pi^3 x^5 },
\label{df-over-dx}
\ee
where we define ${ x={\rm MeV}/T}$.

It is clear from this equation that the freezing temperature, 
${T_f}$, depends upon the neutrino momentum ${ y =p/T}$, and this can
distort the spectrum of the decoupled neutrinos, as we see in what follows.

For the average value of the neutrino momentum, ${ y = 3}$, the 
temperature of decoupling of neutrinos from $e^\pm$ is
\be 
T_{\nue} = 1.87\,\, {\rm MeV},\,\,\,{\rm and}\,\,\,
T_{\num,\nut} = 3.12\,\, {\rm MeV}.
\label{T-nue}
\ee
{\it Problem 12.} Derive equation (\ref{df-over-dx}).
Find decoupling temperature of annihilation, ${\nu\,\bar \nu \lrar e^+e^-}$
which changes the number density of neutrinos. \\
{\it Answer:} ${T_{\nu_e} \approx 3}$ MeV and 
${T_{\nu_\mu,\nu_\tau} \approx 5}$ MeV. 

To take into account all reactions experienced by neutrinos
including elastic ${ \nu e}$ and all ${ \nu \nu}$ scattering we
need to make the substitution: {${(g_L^2 + g_R^2)\rar (1+g_L^2 + g_R^2)}$}
in eq. (\ref{df-over-dx}) and find that
neutrinos started to propagate freely in the
universe when the temperature dropped below 
\be 
T_{\nue} = 1.34\,\,{\rm  MeV} \,\,\,{\rm and}\,\,\,
T_{\num,\nut} = 1.5\,\, {\rm MeV}.
\label{T-free}
\ee

\subsection{Gershtein-Zeldovich(GZ)  bound \label{ss-GZ}}

Consideration of thermal equilibrium and entropy conservation
permitted Gerstein and Zeldovich to derive famous cosmological 
bound on neutrino mass~\cite{gz}. Sometimes this bound is called 
Cowsic-McLelland bound but this is not just because paper~\cite{cow-mcl} 
has been published 6 years after Gerstein and Zeldovich and contained
a couple of inaccurate statements which resulted in overestimation of the 
bound by factor 22/3.

We need to calculate the ratio $n_\nu/n_\gamma$ at the present time. The
known from observations number density of photons in CMB,
see eq. (\ref{n-gamma}), allows to determine the cosmological number
density of unobservable neutrinos. 
At neutrino decoupling the ratio is determined by thermal equilibrium: 
\be
 n_\nu = n_{\bar \nu} = (3/8) n_\gamma
\label{nu-n-gamma-eq}
\ee
After decoupling ${ n_\nu}$ was conserved in the comoving volume,
i.e. {${ n_\nu\,a^3 = const}$} but
${ n_\gamma\,a^3 }$ rises due to  ${ e^+e^-}$--annihilation into photons.
At first sight the rise of $n_\gamma$ is difficult to calculate but entropy
conservation (\ref{dS-dt}) makes the calculations trivial.

After neutrino decoupling the number density of neutrinos fall down
as cosmological volume, ${ n_\nu \sim 1/a^3}$.
Before $e^+ e^-$--annihilation but after neutrino decoupling, say 
at ${ T\sim 1 MeV}$, the entropy of photons and
electron-positron pairs was:
\be
S_{in} \sim (2+7/2) T^3_{in}\, a_{in}^3\,.
\label{S-in}
\ee
After annihilation it became:
\be 
S_{fin} \sim 2 T^3_{fin}\, a_{fin}^3\,.
\label{S-fin}
\ee
Since ${ n_\gamma \sim T^3 }$,  and ${ S_{in} = S_{fin} }$, 
the ratio of number densities of neutrino and photons dropped by the factor
4/11. If in the course of subsequent cosmological evolution the numbers of
photons and neutrinos conserved in the comoving volume, the number density
of neutrinos today must be
\be
n_\nu + n_{\bar\nu} = \frac{3}{11}\,n_\gamma = 112/cm^3
\label{n-nu-today}
\ee 
This result is obtained under assumption of vanishingly small chemical 
potentials of neutrinos, in other words for $n_\nu = n_{\bar \nu}$.

Energy density of neutrinos today should be smaller
than  the total energy density of matter, ${ \rho_m}$. This leads
to the following upper bound on the sum of masses of all neutrino species:
\be 
\sum m_{\nu_j} < 94\,\, {\rm eV}\,\, \Omega h^2\,.
\label{sum-m-nu}
\ee
Since $h^2 \approx 0.5$, $\Omega_m \approx 0.25$, and  
masses of different neutrinos are nearly equal, as follows from the data on
neutrino oscillations, we find $ \mnu < 5$ eV.

This limit may be further strengthen, if one takes into account that 
cosmological structure formation would be 
inhibited at small scales if 
${ \Omega_{HDM}> 0.3\, \Omega_{CDM} }$. Hence {${ \mnu < 1.7\,\, eV}$.}
Recent combined analysis of CMB and LSS leads to the bound:
\be
\mnu < 0.3\,\, {\rm eV}\,.
\label{m-nu-last}
\ee
For more detail and reviews see ref.~\cite{m-nu} 
If the masses of neutrinos are close to this upper limit, their contribution
into cosmological energy density at the present time would be non-negligible,
${ \Omega_\nu \sim 0.02}$,
comparable to  that of baryons, ${\Omega_b \approx 0.04}$.

One may argue that though massless neutrinos are 100\% left-handed, i.e.
they have only one helicity state, massive neutrinos have both spin states
and hence their cosmological number density should be twice larger than 
calculated. However, it is not so because right-handed states did not reach
equilibrium and their contribution may be neglected.

Next question is how robust is the GZ bound. Is it possible to modify the
standard picture to avoid or weaken it. The bound is based on the following
assumptions:
\begin{enumerate}
\item{}
Thermal equilibrium between ${ \nu}$,
${ e^\pm}$, ${ \gamma}$ at ${ T\sim {\rm MeV}}$. If the universe never was
at ${ T\geq {\rm MeV}}$, neutrinos might be under-abundant and the bound would be
much weaker. However, successful description of light element production at BBN
makes it difficult or impossible to eliminate equilibrium neutrinos at the
MeV phase in the universe evolution. 
\item{}
{Negligible lepton asymmetry.} Non-zero lepton asymmetry would result in
larger number/energy density of neutrinos plus antineutrinos and the bound
would be stronger. 
\item{}
{No extra production of CMB photons after neutrino decoupling.}
Strictly speaking this is not excluded but strongly constrained. If the 
extra photons were created before BBN terminated, they might distort 
abundances of light elements. Late time creation of extra photons, after BBN,
would lead to distortion of the energy spectrum of CMB and there is only very 
small freedom, not sufficient to change $n_\nu/n_\gamma$ essentially.     
\item{}
Neutrino stability on the cosmological scale, ${ \tau_\nu> t_U}$. If neutrino
decays into another normal neutrino, e.g. $\mu_{\mu} \rar \nu_e + X$, the total
number of neutrinos does not change and the limit on the mass of the lighest neutrino
remains undisturbed, but heavier neutrinos are allowed.
If the decay goes into a new lighter fermions, e.g. sterile neutrino, the bound may 
be weakened for all neutrino species.
\item{}
No late-time annihilation of ${ \nu + \bar\nu}$ into a pair of (pseudo)goldstone
bosons, e.g. majorons. For noticeable annihilation too strong coupling of 
neutrinos to majorons is necessary which is probably excluded by astrophysics.
\end{enumerate}

\subsection{Distortion of neutrino spectrum \label{ss-nu-sperct} }
As we mentioned above, see eqs. (\ref{Hxdxf}-\ref{lhsm-zero}),
massless particles keep their equilibrium spectrum even after the
interaction is switched off. 
E.g. the spectrum of CMB photons
is the equilibrium one with the precision better than ${10^{-4}}$.
However, it happened not to be true for neutrinos. The point is that 
neutrino decoupling is not instantaneous and for some time there coexist
two components of plasma with different temperatures, weakly interacting with
each other. Indeed, due to ${e^+e^-}$ annihilation the photon temperatures
rises with respect to the neutrino temperature as,
${T_\gamma /T_\nu} \approx 1.4$. 

Due to residual interaction of neutrinos with hotter electrons and positrons
some energy is transferred to colder neutrino sector. 
More energetic neutrinos decoupled later. As a result the spectrum is
distorted~\cite{ad-mf}:
\be 
\delta f_{\nu_e} /f{eq} \approx 3\times 10^{-4}\,\frac{E}{T}
\left(\frac{11E}{4T} -3\right) 
\label{delta-f-nu}
\ee
This analytical estimate was confirmed by precise numerical solution
of the integro-differential kinetic equation; for discussion and
the list of references see review~\cite{nu-rev}.

This effect leads to an increase of the effective number of neutrino species: 
\be
\Delta N_\nu = 0.03 + 0.01.
\label{Delta-N}
\ee
The last 0.01 comes from plasma corrections~\cite{nu-plasma}, which
diminish ${ n_e}$ and ${ n_\gamma}$ with respect to unperturbed
quantities at the same temperature.
An increase of the number of neutrino species
{has negligible effect on BBN, ${ \sim 10^{-4}}$,}
{but may be noticeable in CMB measurements by the recently launched Planck
mission.} 

Note in conclusion of this section that though we mentioned above that
neutrino temperature is approximately 1.4 times smaller than the temperature 
of photons, i.e. today it should be 1.95 K, would neutrino be massless, the
distribution of neutrinos has the non-equilibrium form:
\be
f_\nu \approx [\exp (p/T_\nu) +1 ]^{-1}\,,
\label{f-nu}
\ee
i.e the magnitude of neutrino momentum enters instead of energy and so the 
parameter $T_\nu$ does not have meaning of temperature. The 
correction (\ref{delta-f-nu}) is neglected here.

\subsection{Non-relativistic freezing \label{ss-non-rel-frees}}
 
If particles have sufficiently strong interactions, they decouple from
primordial plasma at temperatures much smaller than their mass,
${ T_f < m_h}$. After that
their number density stopped falling down according the the Boltzmann suppression
law but remains constant in the comoving volume. The number density of heavy
particles at decoupling is given by
\be
n_h/n_\gamma \approx (m_h/T_f)^{3/2} e^{-m_h/T_f} \ll 1,
\label{n-h-n-gamma}
\ee
so such particles may have masses much larger than permitted by GZ bound
and can make cosmological interesting cold dark matter.
The frozen number density of such particles is determined by the cross-section
of their annihilation and is given by a simple expression, see e.g.~\cite{dz}:
\be
\frac{ n_{h}}{{n_{\gamma}}} \approx  \frac{(m_h/T_f)} 
{\langle { \sigma_{ann} v} \rangle m_{Pl} m_h }\,,
\label{nuhnug}
\ee
where ${ m_h/T_f \approx {ln} (\langle 
{\sigma_{ann}} v \rangle m_{Pl} m_h) \sim  (10-50)}$.

We derive this expression in what follows. One can solve
kinetic equation governing evolution of the number density of heavy
particles numerically but it is instructive to make analytic calculations.
Moreover, the results are pretty accurate. 
Analytic calculations of frozen abundances are usually
done under the following assumptions:
\begin{enumerate}
\item{}
Boltzmann statistics. It is usually a good approximation for 
heavy particles at ${ T<m}$.
\item{}
It is assumed that heavy particles are in kinetic, but
not chemical, equilibrium, i.e. their distribution function has the form:
\be
 f_h = e^{-E/T + \xi(t)}\,,
\label{f-h}
\ee
where $\xi$ is the effective chemical potential normalised to temperature,
$\xi = \mu/T$. Chemical equilibrium is enforced by 
annihilation which needs a partner whose number density is 
exponentially suppressed, while kinetic equilibrium 
demands encounter with abundant massless particles. That's why chemical
equilibrium stopped to be maintained much earlier than the kinetic one.
\item{}
{The  products of annihilation are in complete thermal equilibrium.}
\item{}
Charge asymmetry of heavy particles is negligible and thus the chemical
potentials for particles and antiparticles are equal,
{${ \xi=+\bar \xi}$}. Annihilation would be 
much more efficient in the case of non-zero charge asymmetry. 
\end{enumerate}
Kinetic equation under this assumption becomes an ordinary differential
equation, which has been derived in 1965 by Zeldovich~\cite{zel-65} 
and used for the calculation of the frozen number density of non-confined
massive quarks in ref.~\cite{zop}.
In 1978 the equation was applied to the calculations of the frozen number
densities of stable heavy leptons in ref.~\cite{lw, vdz} and after that
it got the name Lee-Weinberg equation, though it would be more proper to call
it Zeldovich equation.

The equation has the following simple form:
\be
\dot n_h +3H n_h = \langle \sigma_{ann} v \rangle
(n^{(eq)2}_{h} -n^2_h)\,,
\label{zel-eq}
\ee
where $n_h$ is the number density of heavy particles, $n_h^{(eq)}$
is its equilibrium value, and $ \langle \sigma_{ann} v \rangle $ is thermally
averaged annihilation cross-section multiplied by velocity of the annihilating
particles:
\be
\langle {\sigma_{ann}} v \rangle =
\frac{(2\pi)^4}{(n_{h}^{eq})^2}
\int{\overline {dp}}_{h}  {\overline {dp}}_{\bar h}
 \int{\overline {dp}}_{f}  {\overline {dp}}_{f'}
\delta^4 (P_{in}-P_{fin})
 |A_{ann}|^2 e^{-\left( E_p  + E_{p'}\right) /T}\,,
\label{sigma-v}
\ee
where ${{\overline {dp}}= d^3p /[2E\,(2\pi)^3]}$.

The integration in eq. (\ref{sigma-v}) can be  taken down to one variable
and we find~\cite{gelm-gond}:
\be
\langle \sigma_{ann} v \rangle = \frac{x}{8 m_h^5 K_2^2(x)}
\int_{4 m_h^2}^{\infty}ds~(s-4 m_h^2)
\sigma_{ann}(s)
\sqrt{s} K_1\left( \frac{x\sqrt{s}}{m_h}\right)
\label{sigma0v2}
\ee
where ${ x = m_h/T}$ and ${ s=(p+\bar p)^2}$.
Usually ${x\gg 1}$ and ${\sigma_{ann} v \rar const}$ near threshold,
so thermally averaged $\langle \sigma v \rangle$ is reduced just to 
the threshold value of $\sigma v$. The expression above can be useful if
cross-section noticeably changes near threshold, e.g. in the case of 
resonance annihilation.

For derivation of eq. (\ref{zel-eq}) we start with the general kinetic
equation:
\be 
\partial_t f - Hp\partial_p f = I_{el} + I_{ann}\,,
\label{gen-kin-eq}
\ee
where we take into account only two-body processes with heavy particles,
$I_{el}$ and $I_{ann}$ are respectively collision integrals for elastic 
scattering and annihilation. At $T< m_h$ the former is much larger than the
latter because of exponential suppression of the number density of heavy
particles, $f_h \sim \exp (-m_h/T)$.  Since ${I_{el}} $ is large, it enforces
kinetic equilibrium, i.e. canonical distribution over energy: 
\be
f_h = \exp [-E/T+\xi(t)].
\label{f-kin-eq}
\ee
With such a form of $f_h$ we can integrate both sides of eq. (\ref{gen-kin-eq})
over ${\bar{dp}}$ and the large elastic collision 
integral disappears, but the trace of it remains in the 
distribution~(\ref{f-kin-eq}). 

As the last step we express ${\xi(t)}$ through ${n_h}$:
\be
\exp (\xi) = n_h/n_{eq}
\label{exp=xi}
\ee
and arrive to eq. (\ref{zel-eq}).

The equation can be solved analytically, approximately but quite accurately. 
Usually at high temperatures, ${T \geq m_h}$, the annihilation rate is high:   
\be 
\sigma_{ann} n_h /H \gg 1
\label{sigma-over-H}
\ee
and thus the equilibrium with respect to annihilation is maintained,
${n_h = n_{eq} +\delta n}$, where $\delta n $ is small. It is convenient to  
introduce dimensionless ratio of number density to entropy 
$ {r= n_h/S}$, so the effects of expansion disappear from the equation: 
\be 
\dot n +3 Hn = S \dot r\,.
\label{S-dot-r}
\ee   
Recall that $S$ is conserved in comoving volume, eq. (\ref{dS-dt}).

By assumption $r$ weakly deviated from equilibrium, so we can write
${ r = r_{eq}(1 +\delta r)}$, where $\delta r \ll 1$. In this limit
the solution of eq. (\ref{zel-eq}) can be found in stationary point 
approximation:
\be
\delta r \approx - \frac{Hx r'_{eq}}{2\sigma v S\,r_{eq}^2}
\label{delta-r}
\ee
Since $r_{eq}$ exponentially drops down, $\delta r$ rises and
at some moment $\delta r$ 
would reach unity. After that we will use another approximation, neglecting
${r^2_{eq}}$, integrate equation for $r$ and obtain the final 
result (\ref{nuhnug}).

Another way to solve equation (\ref{zel-eq}) is to transform this Ricatti type
equation to the second order Schroedinger type one and to integrate the
latter in quasi-classical approximation.  \\[2mm]
{\it Problem 13.}
Derive all above, in particular kinetic equation and solve it.\\
{\it Problem 14.} Find frozen number densities of 
protons and electrons in charge symmetric universe.}
{\it Answer:} ${n_p/n_\gamma \approx 10^{-19}}$, 
${n_e/n_\gamma \approx 10^{-16}}$.\\
{\it Problem 15.} What number density would have anti-protons if
$(n_p-n_{\bar p})/n_\gamma = 10^{-9}$. 

Let us apply the obtained results for calculation of the frozen number density
of lightest supersymmetric particle (LSP) which must be stable if R-parity is
conserved and is a popular candidate for dark matter. 
The annihilation cross-section is estimated as:
\be
\sigma v \sim \alpha^2/m^2_{S}
\label{sigma-ann-LSP}
\ee
Correspondingly the energy density of LSP would be:
\be
\rho_{SUSY} = m_{S} n_{S} \approx \frac{n_\gamma\, m_S^2\, 
\ln(\alpha^2 M_{Pl}/m_S) } {M_{Pl}}
\label{rho-SUSY}
\ee
For ${m_S =100}$ GeV, which is a reasonable value for minimal supersymmetric 
model, we find:
\be
\Omega_{SUSY} \approx 0.05\,.
\label{Omega-SUS}
\ee 
It is very close to the observed 0.25 and makes LSP a
{natural candidate for DM.}

Another interesting example is the frozen number density of magnetic monopoles,
which may exist in spontaneously broken gauge theories containing $O(3)$
subgroup~\cite{magn-mon}. The cross-section of the monopole-antimonopole
annihilation can be estimated as 
\be
{\sigma_{ann} v\sim g^2/M_M^2}, 
\label{sigma-mon}
\ee
where $M_M$ is the monopole mass. Correspondingly the present day energy density
of magnetic monopoles would be~\cite{zel-khlop}:
\be 
\rho_M = \frac{n_\gamma M_M^2}{g^2 M_{Pl}}\,. 
\label{rho-M}
\ee
In fact slow diffusion of monopoles in cosmic plasma would slightly 
diminish the result but not much.

If  ${M_M \sim 10^{17}}$ GeV, as predicted by grand unified theories,
the monopoles would overclose the universe by about 24 orders of magnitude
assuming that their initial abundance was close to the thermal equilibrium one. 
This problem played a driving role for the suggestion of inflationary 
cosmology.

\section{Big bang nucleosynthesis \label{s-bbn}}

BBN is one of the pillars of the standard cosmological model. 
It describes creation of light elements, $^2H$, $^3He$ $^4He$, and $^7Li$, 
in the early universe when she was
between 1 sec to 200 sec old and the temperature ran in the interval from 1 MeV
down to 60-70 keV. The calculated abundances of light elements are in a good
agreement with the observation. This proves that our understanding of the universe
when it was so young, is basically correct.

The first stage of BBN is the freezing of the neutron-to-proton ratio, which 
determines the number density of neutrons for the second phase when formation of 
light elements took place. The $n/p$--freezing happened at 
${T\approx 1}$ MeV and ${t \approx 1}$ s, while the light element formation
occurred much later at ${T\approx 65}$ keV and  ${t \approx 200}$ s. 

The neutron-to-proton ratio is determined by the reactions:
\be
n + e^+ &\lrar& p + \bar\nu_e
\label{n-e}\\
 n + \nu_e &\lrar& p + e^-\,,
\label{n-nu-e}
\ee
which frozen  at ${ T\approx 0.7}$ MeV (see below). After that $r_{np}$ remained
almost constant, slowly decreasing due to the neutron decay:
\be
n  \lrar p + e^- + \bar\nu_e\,,
\label{n-decay}
\ee
whose life-time is $\tau_n = 886 $ s. Since the formation of light elements started
at $t\approx 200 $ s, the decrease of $r_{np}$ due to decay was
essential. However, the decay is not important 
for $(n-p)$ freezing. 

It is more convenient to consider the ratio of neutron number density
normalised to total baryon number density, $r = n_n/(n_p+n_n)$, because latter 
is conserved in comoving volume at BBN epoch since baryonic number was 
conserved at low temperatures.

Kinetic equation which governs the neutron to baryon ratio
can be obtained from the general kinetic equation (\ref{kin-eq}) with collision
integral given by eq. (\ref{I-coll}) in the limit
of non-relativistic nucleons:
\be 
\dot r = {(1+3g^2_A) G_F^2 \over 2\pi^3} \left[ A - (A+B)\,r  \right]
\label{dot-r}
\ee
where ${ g_A = -1.267}$ is the axial coupling constant of $(n-p)$ weak current and
the coefficients $A$ and $B$ are
\be
A &=& \int^\infty_0 dE_\nu K f_e(E_e)
\left[1-f_\nu(E_\nu) \right] |_{E_e = E_\nu + \Delta m} 
 +\int_{m_e}^\infty dE_e K 
f_{\bar \nu} (E_\nu)
\left[1-f_{\bar e}(E_e \right] \mid_{E_\nu = E_e + \Delta m} 
\nonumber \\
&+& \int_{m_e}^{\Delta m} dE_e K 
f_{\bar \nu} (E_\nu)
f_e(E_e)\mid_{E_\nu + E_e = \Delta m}\,,
\nonumber \\
B &=& \int^\infty_0 dE_\nu K f_\nu(E_\nu)
\left[1-f_e(E_e) \right]|_{E_e = E_\nu + \Delta m} 
{{ +\int_{m_e}^\infty dE_e K
f_{\bar e}(E_e) 
\left[1-f_{\bar \nu}(E_\nu) \right]
\mid_{E_\nu = E_e + \Delta m} 
}}\nonumber \\
&+&\int_{m_e}^{\Delta m} dE_e K
\left[1-f_{\bar \nu}(E_\nu) \right]
\left[1-f_e(E_e) \right] \mid_{E_\nu + E_e = \Delta m}\,.
\label{AB}
\ee
where ${ K= E_\nu^2 E_e p_e }$ and we included terms describing neutron
decay, the last ones in expressions for $A$ and $B$.
{In precision calculations relativistic corrections and
all form-factors of $(n-p)$--transformations are taken into account.}

The expressions for $A$ and $B$ take very simple form if $e^\pm$ and $\nu_e$
are in thermal equilibrium with equal temperatures. In this case
{${ A=B\,\exp\left( -\Delta m/T\right)}$.} Moreover, 
in essential range of temperature ${ m_e}$ can be neglected and: 
\be
B = 48T^5 + 24(\Delta m) T^4 + 4(\Delta m)^2 T^3.
\label{B-appr}
\ee
In this approximation equation (\ref{dot-r}) can be easily solved 
numerically.

We can make an estimate of the freezing temperature using the following
simple considerations. The reaction rate versus Hubble rate is:
\be
\frac{\Gamma_{np}}{H} = \frac{(1+3g^2_A) G_F^2 B /2\pi^3}
{ T^2 \sqrt {g_*} /0.6 m_{Pl}}
\label{Gamma-np-H}
\ee 
The ${ n/p}$ freezing temperature can be approximately found from the
condition $\Gamma_{np}/H = 1$, that is
\be
T_{np} = 0.7 \left( {g_*\over 10.75} \right)^{1/6} {\rm MeV}
\label{Tnp}
\ee
and correspondingly
{${ (n/p)_f =\exp(-\Delta m/T_{np})\approx 0.135}$.} Note that 
$ T_{np}$ depends upon the number of neutrino species through
${ g_*}$, see discussion after eq. (\ref{H-of-x}). 

When ${ T}$ drops down to $ T_{BBN} =60-70$ keV, practically
all neutrons quickly form ${^4 He}$ (about
25\% by mass), ${ ^2H}$ (${ 3\times 10^{-5}}$ by number),
${^3 He}$
(similar to ${ ^2H}$), and ${ ^7 Li}$ (${ 10^{-9}-10^{-10}}$). The calculated 
abundances {span 9 orders of magnitude and well agree with the data.}\\
{\it Question:} why ${ T_{BBN}}$ is much smaller than nuclear
binding energy, $E_b \sim {\rm MeV}$? The answer is below in this section.

Almost all frozen neutrons, except for those
which decayed before the onset of light element formation at 
{$ { T= T_{BBN} \approx 65}$ keV,}
form ${ ^4 He}$ because
of its largest binding energy equal to 7 MeV/nucleon.
Correspondingly the mass fraction of  ${ ^4 He}$ can be estimated as: 
\be
Y = 2(n/p)/[1+(n/p)] \approx  24\%.
\label{Y} 
\ee
in a good agreement with the data.

It is interesting that a small variation of the Fermi coupling constant
would strongly change the amount of the produced $^4 He$ and correspondingly the
star properties. The stars might either have a deficit of helium or of hydrogen.\\
{\it Problem 16.} Find the range of variation of ${ G_F}$ which is in agreement with
${ (25\pm 1)\%} $ of the mass fraction of ${ ^4 He}$.
Find the same for ${ g_*}$ or the number of neutrino families.

The light element formation proceeded through the chain of reactions:
{${ p\,(n,\gamma)\,d}$, ${ d\,(p\gamma)\,^3He}$, 
${ d\,(d,n)\,^3He}$, ${ d\,(d,p)\,t}$,
${ t\,(d,n)\,^4He}$, {\it etc}.} All reaction
go through formation of deuterium, because due to low baryon
density two body processes dominate.
An absence of a stable nuclei with ${ A=5}$ results in suppression of 
heavier nuclei production.

Let us turn now to the calculations of the temperature when the light elements
were created. Naively one should expect this temperature to be close to the
nucleus binding energy $T\sim E_b$. However, a large number of photons make 
it possible to destroy the produced nuclei on the tail of their energy distribution,
despite the Boltzmann suppression. So one would expect that the temperature 
of nucleus formation is smaller than the binding energy by the logarithm of the
ratio $n_B/n_\gamma$. We will derive now the Saha equation and
see that this is indeed that case. 

The equilibrium density of deuterium is determined by the equality
of chemical potentials, ${ \mu_d = \mu_p + \mu_n}$:
\be
n_D = 3\left(\frac{m_D T}{2\pi}\right)^{3/2} 
e^{\left[(-m_D+\mu_p+\mu_n )/T\right]}\,,
\label{n-D-eq}
\ee
where the chemical potential can be expressed through the number density as  
\be
e^{\mu_p/T} = \frac{1}{2}\, n_p\, 
\left(\frac{2\pi}{m_p T}\right)^{3/2} e^{m_p/T}
\label{mu-p}
\ee
Thus in equilibrium we can express the number density of deuterium through the 
number densities of protons and neutrons:
\be 
n_d =
{3\over 4}n_n n_p e^{ B_D/T } \left( 2\pi m_d \over m_pm_n T \right)^{3/2}\,,
\label{nd}
\ee
where ${ B_D = m_p+m_n-m_D= 2.224} $ MeV is the binding energy of deuterium 
and ${ n_p =\eta n_\gamma}$ with ${{ \eta \equiv \beta = 6\cdot 10^{-10}\ll 1}}$.
Notation $\eta$ is used at consideration of BBN, while the same or 
similar quantity is 
denoted as $\beta$ when baryogenesis is considered.

The number density of deuterium, ${ n_d}$, becomes comparable to ${ n_n}$ at
\be
T_{BBN} {\approx}
\frac{B_D}{\ln(16/3 \eta) + 1.5 \ln (m_p/4\pi T)}=
\frac{ 0.064 \,{\rm MeV}}{ 1 - 0.029\, \ln \eta_{10}}\,,
\label{T-BBN}
\ee
where $\eta_{10} = 10^{10}\eta$. At higher temperatures $n_D$ (and densities of other 
nuclei) are much smaller than $n_n$ and thus $T_{BBN}$ can be considered as the 
temperature when the formation of light nuclei started. Because of the exponential
dependence on $T$ the light nuclei formation proceeded during quite short time interval.

Note, that due to the same effect, the
hydrogen recombination took place at $T\sim 0.1$ eV which is much smaller
than the hydrogen binding energy, $E_H = 14.6$ eV.

\begin{figure}
\begin{center}
\includegraphics[height=0.7\textheight]{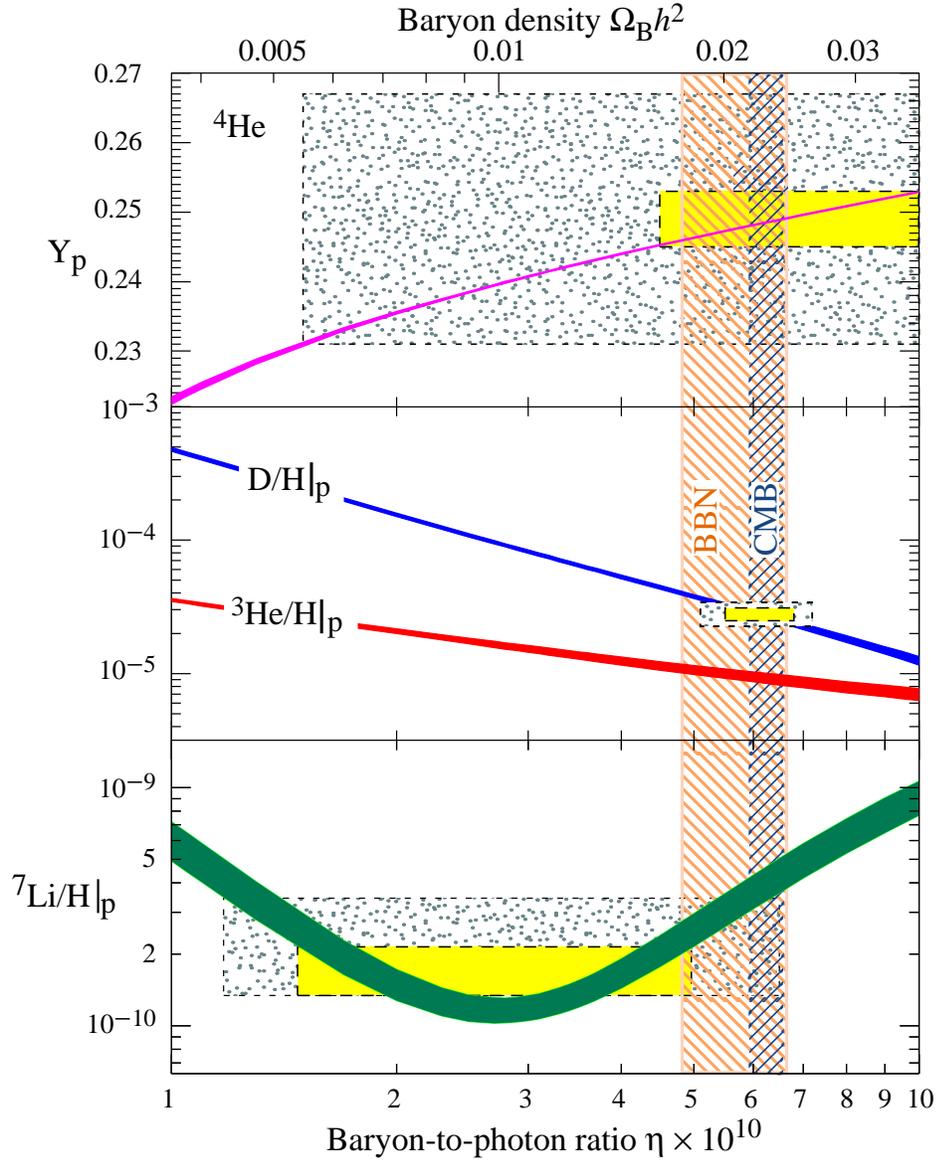}
\caption{
The abundances of He$^4$, D, He$^3$ and Li$^7$ 
predicted by the standard model of BBN.
The bands show the 95\% CL range. Boxes
indicate the observed light element abundances (smaller boxes: $\pm 2\sigma$ 
statistical errors; larger boxes: $\pm 2\sigma$
statistical and systematic errors). The narrow vertical
band indicates the CMB measure of the cosmic baryon density, while the wider
band indicates the BBN concordance range (both at 95\% CL).
} \label{bbn-obs}
\end{center}
\end{figure}

In Fig. \ref{bbn-obs}, taken from ref.~\cite{pdg}, 
the calculated abundances of light elements as functions
of the baryon-to-photon ratio are presented. 

Helium-4 slowly rises with ${\eta}$, because for larger $\eta$
the moment of BBN becomes earlier, see eq. (\ref{T-BBN}), 
and less neutrons decayed.

Deuterium abundance quickly drops down with rising ${\eta}$ because the probability 
of processing of deuterium to heavier more tightly bound nuclei, $^4He$,
is larger with larger baryonic number density. So less deuterium survives at
larger $\eta$. High sensitivity of the
primordial deuterium abundance to $\eta$ allowed it to be the best way to 
measure the cosmological amount of baryons before more accurate CMB
measurements became available. This is why deuterium was called ``baryometer''.

Lithium-7 is formed in two competing processes with different dependence on
${\eta}$. At low $\eta_{10} < 3$ the production predominantly goes through the
reaction $^3H (^4He,\gamma) ^7Li$. On the other hand, $^7Li$ can be destroyed by 
collisions with protons and with rising $\eta$ destruction becomes more efficient
and $^7Li$ drops down. At larger $\eta_{10} > 3$ the dominant process of creation
is $^3He (^4He,\gamma)^7Be$. Destruction of $^7Be$ by protons is less efficient 
because $^7Be$ has larger binding energy than $^7Li$. So $^7Be$ production rises 
with rising $\eta$. At lower $T$, when formation of atoms became non-negigible, 
$^7Be$ could capture electron and decay into $^7Li$ and neutrino.

The BBN calculations are based on pretty well known low energy 
nuclear physics and theoretical uncertainties would not play a significant
role in comparison of theory with observations, if we were able to observe
these light elements at the epoch of their creation.
However, we observe them now, while the results are obtained for very young
universe, about 300 seconds old. So evolutionary effects should be taken into 
account. We will briefly describe the problems of comparison of theory with the
data below. For more detailed discussion see reviews~\cite{pdg,steigman}. 

Helium-4 is very tightly bound nuclei and so it is
not destroyed in the course of cosmological evolution. Hence  
the observed abundance of $^4He$ should be larger than the primordial one.
With the existing observation means $^4He$ can be observed only at low red-shifts in
chemically evolved regions with the abundance which may be quite different from the
primordial one. To deduce the primordial abundance of $^4He$ one needs to 
extrapolate to zero metallically. Namely the regions where $^4He$ is observed
are contaminated by heavier elements. One can study the correlation of this
elements with $^4He$ and extrapolate (linearly) the data to zero values of the
metals (in astronomy all heavier than helium are called metals). Another source 
of uncertainty is not very well known fraction of ionised helium with respect to the
total amount. This described by the so called ionisation correction, which is another
source of uncertainty.

Deuterium could be destroyed in the course of evolution in poorly controlled
manner. Fortunately, in contrast to helium, the deuterium line can be observed
at large red-shifts, $z\sim 1$, i.e. at the earlier stages of the cosmological evolution.
If the clouds, where deuterium is observed, are not contaminated by heavier elements,
there is a good chance that deuterium is primordial. However, the deuterium line is 
shifted from the hydrogen one only by 80 km/sec and the peculiar motion of the cloud
may induce an essential systematic error. The average value of deuterium abundance
is in good agreement with the value of $\eta$ determined from CMB, but the individual
values are rather strongly dispersed from ${1.6\cdot 10^{-5}}$
up to ${3.5\cdot 10^{-5}}$. It would be interesting to understand the origin of such
strong dispersion.

Primordial lithium visibly creates a potential problem for BBN, but $^7Li$ is
difficult to observe (in first generation stars) and maybe it is premature to worry
about the disagreement.

As a whole the data and theory are in a good agreement. Still
there seems to be some ``small clouds''. It would be very interesting if these clouds
indicate new physics but most probably the resolution of the problems can be
found in the traditional way when more accurate data are accumulated and better 
understood.

\section{Role of neutrinos in BBN \label{s-nu-bbn}}

BBN is sensitive to any form of energy which was present in the universe during
light element formation. Indeed the universe cooling rate can be determined by 
equating two expressions for the cosmological energy density, namely the critical
energy density and energy density of relativistic plasma with temperature $T$:
\be 
\rho = \frac{3m_{Pl}^2}{32\pi t^2} = \frac{\pi^2}{30}\,g_* T^4\,, 
\label{rho-c-rho-of-T}
\ee
compare to eq.~(\ref{H-of-x}). The factor $g_* = 10.75 + 1.75 \Delta N_\nu$  
count the contributions from photons, $e^\pm$, neutrinos, and any other
form of energy parametrized by ${ \Delta N_\nu}$. Usually $\Delta N_\nu$ is
called the number of extra neutrinos, though the corresponding additional energy may 
have nothing to do with neutrinos. If $\Delta N_\nu =1$, the additional energy is
equal to the equilibrium energy density of one family of neutrinos plus antineutrinos 
with negligible mass (at BBN). In particular, if neutrinos are degenerate, i.e. their
chemical potential $\mu$ is non-zero, the additional energy density corresponds to
\be
\Delta N_\nu = \frac{15}{7}\left[\left(\frac{\xi}{\pi}\right)^4+
2\left(\frac{\xi}{\pi}\right)^2\right]\,,
\label{Delta-N-nu}
\ee
where ${ \xi =\mu/T}$.

{\it Problem 17}. Derive eq. (\ref{Delta-N-nu}). Use eq. (\ref{rhoeq}) and, if necessary,
consult e.g. book~\cite{ll5}, chapter V, sec. 58
for the calculation of the integral. 

There are two effects induced by variation of $g_*$.
First, larger ${\Delta N_\nu}$ leads to earlier ${n/p}$-freezing and higher ${n/p}$-ratio,
see eq. (\ref{Tnp}). Second, with larger $g_*$ the BBN temperature (\ref{T-BBN})
would be reached faster and more neutrons could survive against decay prior the onset
of the light element formation. 
Both effects work in the same direction and $\Delta N_\nu =1$
would lead to an increase of ${ ^4 He}$ by 5\%. Depending upon the data analysis the 
existing observational limit is
\be
\Delta N_\nu < 0.3 - 0.5.
\label{Felta-N-bound}
\ee
According to ref.~\cite{steigman}, ${ N_\nu \approx 2.5}$ seems to be the best fit.
What is it, a problem or an ``experimental'' error?

Extra energy in electronic neutrinos have an additional and stronger effect on BBN
because they can shift the equilibrium value of ${ n/p}$-ratio:
\be{
\left( n/p \right)_{eq} = \exp \left( -\frac{\delta m}{T} -\xi_{e}
\right)
}\label{nu-e-n-p-eq}
\ee
Hence the bounds on lepton asymmetries depend upon the neutrino flavour and are much
more restrictive for $\nu_e$, than for $\nu_{\mu,\tau}$:
\be
|\xi_{\mu,\tau}| < 2.5,\,\,\,\,
{{ |\xi_{e}| < 0.1,}}
\label{xi-mu-xi-e}
\ee
if a compensation between effects induced by ${{ \xi_{\mu,\tau}}}$ and 
${ \xi_e}$ is allowed. 
In absence of compensation the bounds are somewhat stronger.
However, this results were obtained in the case of weak mixing between different
neutrino flavours. In real case of large mixing angle solution to neutrino anomalies
the bounds on all chemical potentials are equal and quite strong, 
see below, sec.~\ref{s-nu-osc}. 

The results presented above are valid for the equilibrium distributions of neutrinos.
If $\nu_\mu$ or $\nu_\tau$ were out of equilibrium at $(n-p)$--freezing their only
effect would be a change of the energy density or, what is the same a contribution
into $\Delta N_\nu$. As for $\nu_e$, their deviations from equilibrium would 
directly change the freezing temperature, $T_{np}$. The effect depends upon the
distortion of the neutrino energy spectrum. In particular, an excess of $\nu_e$ at
high energies would shift $T_{np}$ to lower values and would lead to a smaller 
$n/p$--ratio, which is opposite to the discussed above effect from an increase of 
the total energy density.

Sensitivity of BBN to additional energy at $T\sim 1 $ MeV is used for deriving
bounds on the number density of new (light) particles or new interactions of, say,
neutrinos. As is discussed above, the parameter $g_*$, which describes the contribution
of different species into cosmological energy density, cannot differ form the canonical 
value 10.75 more than by 1. If neutrinos are massive there should be additional 
right-handed states in addition to the usual left-handed ones. If right-handed 
neutrinos are created by the canonical weak interactions, their production probability
is proportional to $m^2_\nu$. The most favourable period for their production took place
at $T\sim 100$ GeV through decays of $W$ or $Z$ bosons. The probability of 
production of ${ \nu_R}$ is
\be
\Gamma_R =\frac{\dot n_{\nu_R}}{ n_{\nu}} =
10\,\left(\frac{m_\nu}{T}\right)^2 \,
\frac{\Gamma_W^\nu n_W + \gamma_Z^\nu n_Z}{T^3}.
\label{Gamma-R}
\ee
One can check that $\nu_R$ have never been produced abundantly if ${ m_\nu}$ 
respects the GZ-bound.

Right-handed neutrinos could be produced directly if there exist right-handed
current induced e.g. by exchange of right-handed intermediate bosons, ${ W_R}$. 
At some early stage of cosmological evolution $\nu_R$ might be abundantly created
and this would endanger successful predictions of BBN. To avoid this problem 
$\nu_R$ should decouple before the QCD phase transition (p.t.). In this case their
number and energy densities would be diluted by the entropy factor, as e.g. number
density of the usual neutrinos were diluted by $e^+e^-$--annihilation discussed in
sec.~\ref{ss-GZ}. The number of species above the QCD p.t. is 58.25, which includes
three quark families (u,d,s) with three colours and 8 gluons with two spin states
plus the usual 10.75 from $e^\pm$, $\gamma$, and $\nu$.
The energy dilution factor is $(10.75/58.25)^{4/3} \approx 0.105$. That is if three $\nu_R$
had equilibrium energy density before QCD p.t., their energy density at BBN would
make 0.3 of the energy density of the one normal neutrino.

Since the ratio of the production rate of $\nu_R$ to the Hubble
parameter is equal to:
\be
{\Gamma_R}/{H} = \left(T/T_W \right)^3 \, \left( m_M/M_{W_R} \right)^4\,,
\label{Gamma-R-H}
\ee
$\nu_R$ would decouple before QCD p.t. if
\be
\frac{m_{W_R}}{m_{W_L}} > \left(\frac{T_{QCD}}{200 {\rm MeV}}\right)^{4/3}.
\label{M-W-R}
\ee  
Here $T_W$ is the decoupling temperature of the normal neutrinos which is taken to be
3 MeV. Thus $ m_{W_R} >$ (a few) TeV, which is
an order of magnitude better than the direct experimental limit.

When/if the precision in BBN will reach the level $\Delta N_\nu < 0.1-0.2$, the
bound on $m_{W_R}$ would be much stronger than above, namely about $10^4$ TeV,   
because we will need to move to the electroweak phase transition or higher.

Using similar arguments we can find a bound on the mixing 
between ${W_R}$ and ${W_L}$:
\be
W_1 = \cos\theta\, W_L + \sin \theta\, W_R 
\label{W-mix}
\ee  
The probability of production of ${\nu_R}$ at ${ T \leq T_{QCD}}$ 
in this case is given by:
\be
r=\Gamma_R/H = \sin^2\theta \left( \frac{T_{QCD}}{200\,{\rm MeV}}\right)^3\,.  
\label{r-Gamma-R-H}
\ee
The condition ${r<0.3}$ results in ${{ \sin^2 \theta < 10^{-3}},}$ or
smaller. The effect does not vanish for ${m_{W_2} \rar \infty}$.
Due to the presence of $\nu_{e_R}$ the ${(n-p)}$--transformation would be more efficient
and it would shift $T_{np}$ to smaller values, thus compensating the increase of $g_*$,  
but the effect is small, ${ \sim \theta^2}$. 
However, it may open window for very large ${ \theta\sim 1}$.

If neutrino has a non-zero magnetic moment,
${\nu_R}$ would be produced in electromagnetic interactions:
\be 
e^\pm + \nu_L \rar e^\pm + \nu_R
\label{ee-nuR}
\ee
Demanding ${ \Delta N_\nu < 0.5}$ we obtain:
\be 
\mu_\nu < 3\times 10^{-10} \mu_B
\label{mu-nu-1}
\ee
If there existed primordial magnetic field, then $\nu_R$ could be produced by
the spin precession in this field and the bound on the magnetic moment would be:
\be
\mu_\nu < 10^{-6} \mu_B \left(B_{primord}/
{\rm { Gauss}}\right)^{-1}
\label{mu-nu-2}
\ee
If there exist intergalactic magnetic fields with the strength 
${ B_{int-gal} \sim 10^{-6}}$ Gauss and if these fields 
were generated in the early universe and evolved adiabatically, then:
\be
\mu_\nu <  10^{-19} \mu_B\,.
\label{mu-nu-3}
\ee

If there exists mirror matter which is similar or identical to ours, BBN
demands that the temperature of the mirror staff should be smaller than the 
temperature of the usual matter  roughly by factor 2,
see e.g. ref.~\cite{mirror}.   

More detail about the material of this and the next section can be found in 
review~\cite{nu-rev}.

\section{Neutrino oscillations in the early universe \label{s-nu-osc}}

Neutrino oscillations in medium are modified in the same way as light propagation.
It can be described by refraction index or what is the same (up to energy factor) 
by effective potential. In cosmological plasma the effective potential contains
two terms~\cite{raf-sigl}:
\be
V_{eff}^a =
\pm C_1 \eta G_FT^3 + C_2^a \frac{G^2_F T^4 E}{\alpha} ~,
\label{V-eff}
\ee
where ${C_j\sim 1}$ and ${\eta}$ is the plasma charge asymmetry:
\be
\eta^{(e)} &=&
2\eta_{\nu_e} +\eta_{\nu_\mu} + \eta_{\nu_\tau} +\eta_{e}-\eta_{n}/2 \,\,\,
 ( {\rm { for}} \,\, \nu_e) \\
\eta^{(\mu)} &=&
2\eta_{\nu_\mu} +\eta_{\nu_e} + \eta_{\nu_\tau} - \eta_{n}/2\,\,\,
({\rm { for}} \,\, \nu_\mu)
\label{eta-nu}
\ee
Effective potential is proportional to the amplitude of forward elastic scattering
(the same as the optical refraction index) and is usually calculated in the lowest
order in the coupling constant. In our case it is first order in $G_F$.

In the local, 4-fermion, limit 
the Lagrangian describing elastic neutrino interaction has the form
\be
{\cal L}_\nu = (G_F/\sqrt{2}) \sum_f
g_f \bar \psi_\nu \gamma_\alpha (1+\gamma_5) \psi_\nu
\bar\psi_f  \gamma_\alpha (1+\gamma_5) \psi_f\,,
\label{L-nu}
\ee
where $f$ are fermions with which neutrino interacts (they include nucleons, charged 
leptons, and
neutrinos) and $g_f$ is the proper coupling constant. This interaction form can be
read-off any textbook on weak interaction.

The first term comes from the averaging of the current 
$J_\alpha \bar\psi_l  \gamma_\alpha (1+\gamma_5) \psi_l\ $ over medium. Since the
cosmological plasma is assumed to be homogeneous and isotropic, the average value of
the space component of the current vanishes but $\langle J_0\rangle \neq 0$, if
the plasma is charge asymmetric. The result is proportional to the difference of the
number densities of particles and antiparticles. This is the first term in 
eq. (\ref{V-eff}). The second term arises from non-locality of weak interactions
due to $W$ or $Z$ boson exchange. At low energies the effect is proportional to
$q^2/m_{W,Z}^2$, where $q$ is the momentum transfer. The second term looks formally
as being of the second order on the coupling constant, but it is imply because
the intermediate boson mass is written as $m^{-2} \sim G_F /\alpha$, where $\alpha$ is the
fine structure constant. In stellar interior $V_{eff}$ is dominated by the first term,
while in cosmology the second term takes over at $T\geq 10$ MeV.

There is another significant complication in cosmology because in contrast to 
stars (except for supernovae at dense stage), the  of neutrino absorption and
production and of coherence breaking due to scattering are significant.
Because of that the standard wave function description is not adequate, the system is 
essentially open and the density matrix formalism should be applied~\cite{dens-mtrx}.
The equation for density matrix has the following form:
\be{
\dot\rho}={\left( {\partial \over \partial t} -
Hp {\partial \over \partial p} \right) \rho}
&=&{ i\left[ {\cal H}_m + V_{eff}, \rho \right] +} 
{ \int d\tau (\bar \nu, l,\bar l) \left( f_l f_{\bar l} A A^+ -
{1\over 2} \left\{ \rho, A \bar\rho A^+ \right\} \right)+}  \nonumber \\
&& \int d\tau (l,\nu',l') \left( f_{l'} B \rho' B^+ -
{1\over 2} f_l \left\{\rho, B B^+\right\} \right)\,.
\label{decoh1}
\ee
A and B are matrix amplitudes of scattering and annihilation respectively.
The equation looks very complicated but numerical solution is possible. Moreover,
in some cases, e.g. if there is MSW-resonance, even an accurate analytical solution 
can be found.

Let us discuss now how neutrino oscillation may influence BBN bounds on lepton
asymmetry. It seems that for large lepton asymmetry, $L$, the oscillations are 
expected to be inhibited due to large first term in effective potential (\ref{V-eff}). 
This term is non-negligible even if $L\sim 10^{-9}$ and for $L\sim 1$ it might kill the
oscillations at all. This is indeed the case for mixing between active and sterile 
neutrinos. However, for mixed {active neutrinos} off-diagonal terms in effective 
potential stimulate oscillations even in presence of large $L$~\cite{panta}.

If the oscillations are not suppressed, individual leptonic numbers would not be
conserved and e.g. initial asymmetry in, say, muonic sector would be equally
redistributed between all three 
flavours: $L_e$, $L_\mu$, and $L_\tau$. Thus the strict
BBN bound on $L_e$, eq. (\ref{xi-mu-xi-e}), became valid for all three leptonic numbers.
The efficiency of redistribution of initially large lepton asymmetry by the
oscillations was calculated in ref.~\cite{dhpprs}, for more related references and
analytical calculations see review~\cite{nu-rev}. It has been shown that for weak
mixing (LOW solution for the solar neutrino anomaly) the redistribution
of  leptonic numbers is insignificant, as is presented in fig.~2. 
However, for large mixing angle solution the equilibration of all leptonic numbers 
proceeded quite efficiently and they all became equal when the neutron freezing occured,
see fig. 3. As a result, the following quite restrictive bound on asymmetry of
all neutrino flavours can be obtained~\cite{dhpprs}:
\be
|\xi_a | < 0.07.
\label{xi-a}
\ee
If this is the case, neutrino degeneracy cannot have a noticeable 
cosmological impact, in particularly on LSS and CMBR.
The bound can be relaxed if 
neutrinos have new interaction with light 
Majorons. The potential induced by the majoron exchange inhibits early 
oscillations and different lepton numbers do not equalise~\cite{ad-ft}.

\begin{figure}[ht]
\label{flow}
\begin{center}
\hspace{10cm}
\vspace{-1.0cm}
\epsfig{file=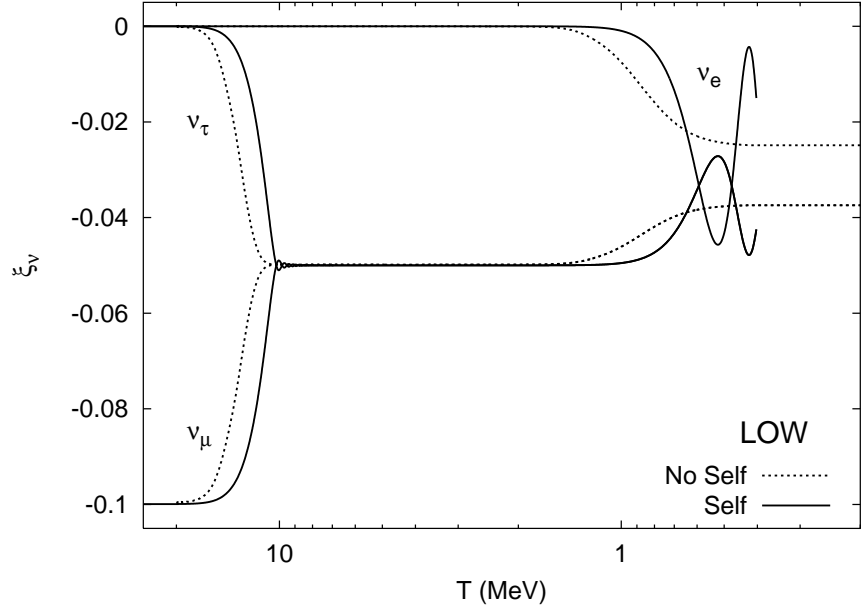,width=12cm}
\vspace{2cm}
\caption{Evolution of lepton charge asymmetry for LOW solution 
to solar anomaly.}
\end{center}
\end{figure}

\begin{figure}[ht]
\begin{center}
\hspace{10cm}
\vspace{-1.0cm}
\epsfig{file=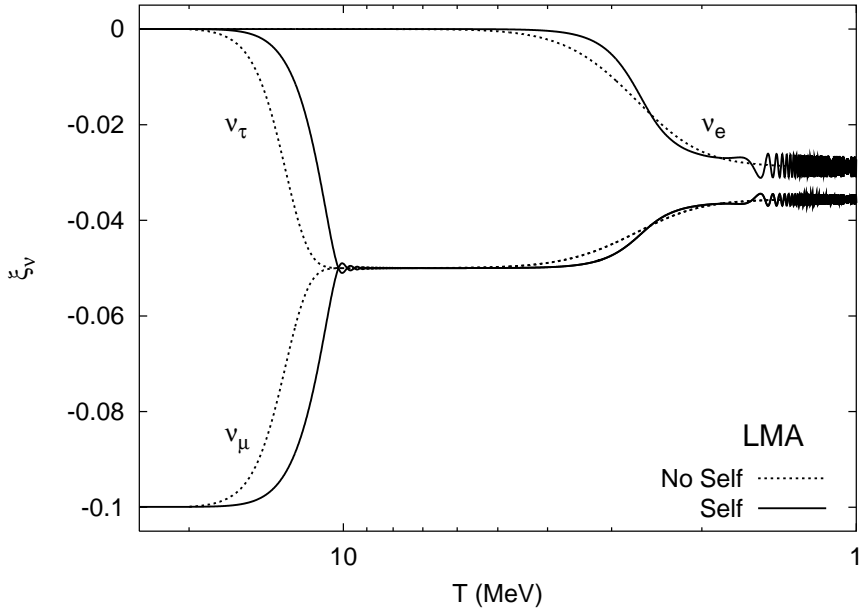,width=12cm}
\vspace{2cm}
\caption{Evolution of lepton charge asymmetry for LMA solution 
to solar anomaly.}
\end{center}
\label{flma}
\end{figure}

If there exists one or several sterile neutrino, $\nu_s$, their mixing with the
usual active ones would lead to several effects potentially observable in BBN:\\
1. An excitation of additional neutrino species leading to positive $\Delta N_\nu$.\\
2. Modification of spectrum of ${ \nu_e}$, because the oscillation probability
depends upon neutrino energy. \\
3. Generation of large lepton asymmetry in $\nu_e$.\\
The impact of these processes on BBN was discussed in a large numbers of works, 
see review~\cite{nu-rev}.
In (non-realistic) case that active nus are not mixed and the mixing between
$\nu_s$ and $\nu_a$ of a certain flavour is not resonance, the following 
bounds can be derived:
\be
(\dm_{\nu_e\nu_s}/{\rm eV}^2) \sin^4 2\theta^{\nu_e\nu_s} &=&
3.16\cdot 10^{-5} \ln^2 (1- \Delta N_\nu)\,,
\label{nu-e-nu-s}\\
(\dm_{\nu_\mu \nu_s}/{\rm eV}^2) \sin^4 2\theta^{\nu_\mu\nu_s} &=&
1.74\cdot 10^{-5} \ln^2 (1-\Delta N_\nu)\,.
\label{nu-mu-nu-s}
\ee
They are noticeably stronger than the bounds obtained from direct experiment.

If the realistic mixing between $\nu_{e,\mu,\tau}$ are taken into account there are
no simple analytical expressions for the BBN bounds, but they have been obtained
numerically in ref.~\cite{ad-fv} both in non-resonance case where they are similar
to those above, eq. (\ref{nu-e-nu-s},\ref{nu-mu-nu-s}), and for resonance case, where
the bounds are much stronger.

\section{Inflation \label{s-infl}}

\subsection{General features \label{ss-gen-inf}}

Inflation is a period of exponential (or more generally accelerated) expansion
of the very early universe, with approximately constant Hubble parameter:
\be
a(t) \sim \exp [ H_i t]\,.
\label{a-exp-Ht}
\ee
It is the earliest time in the universe history about which we can say that it
surely existed.

Inflation is easy to realise e.g. by a scalar field, inflaton, with the 
energy-momentum tensor:
\be
T_{\mu\nu} = \partial_\mu \phi \partial_\mu \phi 
- (1/2)g_{\mu\nu} \left[\partial_\alpha \phi \partial^\alpha \phi
- U(\phi) \right]
\label{T-mu-nu-phi}
\ee
If field $\phi$ slowly evolves, i.e. if 
${ (\partial_\mu \phi)^2 \ll U(\phi)}$, then ${T_{\mu\nu} \sim g_{\mu\nu}}$ and
the vacuum-like equation of state (\ref{T-vac}) would be realised.

The natural question, why do we need such a regime, has a very simple answer: 
inflation is the only {\it known} way to create the observed
universe suitable for life.
Inflation naturally solves previously insolvable problems of Friedman cosmology:\\
1. Flatness. Without inflation the fine-tuning of $|\Omega -1|$ should be
$ 10^{-15}$ at BBN and
$ 10^{-60}$ at the Planck era, see eq. (\ref{Omega-of-t}) 
Otherwise the universe would re-collapse in much less than
${10^{10}} $ years, or expand too fast to make structures.
Since ${|\Omega - 1| \sim \exp (-2H_i t_i)}$, the duration
of inflation should be: 
\be 
H_i t_i > 70\,.
\label{Hi-ti}
\ee
More precisely the minimum duration of inflation may be somewhat smaller depending 
upon the heating temperature after inflation.\\
{2. Causality.}
The distance that CMB photons can propagate in the Friedman universe,
before they stopped interacting, was calculated in sec.~\ref{ss-expan}, 
eq.~\ref{l-gamma}. CMB photons stopped to interact with the surrounding 
medium after the hydrogen recombination at the red-shift
${{ z_{rec}\approx 10^3}}$, see eq. (\ref{T-BBN}) and below. At this moment
the universe age was
${{t \approx 10^{13}}}$ s. Thus it is the maximal size of
the region  ${ d_c }$,
which could be connected by interactions in Friedman cosmology,.
{After recombination ${d_c}$ rises due to the cosmological expansion by
${z_{rec} =10^3}$ and today becomes ${\sim10^{16}}$ s.} 
The angular size of this path in the sky is:
\be
\theta_{max} = \frac{10^{16}}{2\pi\,t_U} \approx 1^o\,,
\label{theta-max}
\ee
and regions outside this size should not know anything about each other.
On the other hand, CMB comes practically the same from all the sky. 
At inflationary stage the causally connected region is exponentially large, so
inflation could make all the observed universe {causally connected if} 
$ H_{i} t_i > 70 $, the same as above.\\
3. Inflation explains the origin of the initial push which induced cosmological 
expansion by antigravity of an almost constant scalar field, since for such a state 
${{P \approx -\rho}}$ and according the eq.~(\ref{ddot-a}) expansion 
speeds up and not slows down as naturally expected with attractive gravity.\\
{4. Inflation makes the universe almost
homogeneous and isotropic at the present-day
Hubble scale.}
Indeed, any perturbation with ${\delta/\rho /\rho \sim 1}$
with wave length ${\lambda}$ would transform into the perturbation with the
same amplitude but with exponentially larger wave length:
\be 
\lambda \rar \exp (H_i t_i) \lambda
\label{lambda-exp}
\ee
In other words, perturbations at fixed scale exponentially 
smooth down.\\
{5. Though inflation kills pre-existing perturbations, it 
creates its own 
small inhomogeneities, at the level $10^{-5}$, but at astronomically large 
scales, which become seeds of
large scale structure (LSS) formation.} 
{Inflation predicts adiabatic Gaussian density perturbations 
with almost flat Harrison-Zeldovich spectrum. In terms of
dimensionless gravitational potential the spectrum does not contain
any dimensional parameter and has approximately a simple form:  
{${ \delta \Psi \sim \delta k/k}$}. 
{Deviations from flat spectrum agree with the data.}

{\it Problem 18}.
According to the existing models, the natural duration of inflation
is much larger than ${H_i t_i = 70}$, hence one should expect
{${ |\Omega -1|}$} negligibly small but in reality  
{inflation predicts at horizon scale today: 
${ |\Omega -1| \sim 10^{-4}}$}. Why?

The idea of inflation was probably the most important breakthrough in cosmology
of the XX century after the big bang one. 
Historically first paper where exponential expansion was invoked for solution
of some problems of Friedman cosmology was that by Kazanas~\cite{kazanas}, who
suggested that exponential regime could solve the problem of the observed
homogeneity and isotropy of the universe. A few months later a famous paper by
Guth ``Inflationary universe: A possible solution to the horizon and flatness 
problems'' was published~\cite{guth}. This work has initiated a stream of papers
which remains unabated to the present day.  
In both scenarios the vacuum-like energy, which might dominate 
at first order phase transition, was suggested as a driving force of exponential
expansion. It was soon understood that such mechanism was not satisfactory
because it would create an inhomogeneous 
universe consisting of many relatively small bubbles in exponentially expanding 
vacuum-like background. The first workable mechanism of inflation was suggested
by Linde~\cite{linde-new} and Abrecht and Steinhardt~\cite{alb-stein}. Probably
the most beautiful inflationary mechanism, the so called chaotic inflation, was 
proposed by Linde~\cite{linde-chaot}. For a review on these and discussed below
issues, see refs.~\cite{inf-rev}. 

There was significant ``pre-inflationary'' literature directly related to the
subject. The idea that the universe avoided singularity and
underwent exponential period during which
the mass of the cosmological matter rose by tens orders of magnitude was discussed 
by Gliner~\cite{gliner} and Gliner and Dymnikova~\cite{gl-dym}.
De Sitter like (exponentially expanding) non-singular cosmology was considered
by Gurovich and Starobinsly~\cite{gur-star} and by Starobinsky~\cite{star}.  
In the last paper an important result was obtained that during ``initial''
exponentially expanding stage gravitational waves were produced which  
may be observable at the present time. If observed, it would be
one of the strongest ``experimental'' indications to primordial inflation.

Another prediction of inflation is the spectrum of primordial density perturbations
which is already verified by the data. The pioneering calculations of the
spectrum have been done by Mukhanov and Chibisov~\cite{mukh-chib} and
confirmed by many subsequent studies~\cite{inf-rev}.

It was shown in the paper by Sato~\cite{sato-1pt} that exponential expansion
induced by first order phase transition would never be terminated for certain 
under-critical values of the parameters. This happened to be a serious shortcoming of 
suggested later first inflationary scenarios. It was also notice by Sato~\cite{sato-anti} 
that exponential expansion might permit astronomically interesting antimatter domains.

\subsection{Models of inflation \label{ss-models}}

There are now many dozens of different mechanisms of inflation, see reviews~\cite{inf-rev},
and we do not have time to talk about them. We will discuss here only one which looks 
most economic and simple; it originated from the Linde's suggestion of
chaotic inflation~\cite{linde-chaot}.

Let us assume that there existed in the very early universe a small piece where
some scalar field (we call it inflaton) is almost constant. In other words
all derivatives, ${\partial \phi} $ are small. We will see that this piece of the
universe would expand exponentially and spatial derivatives would smooth further down. 
The size of this smooth part of the universe should be larger than the inverse Hubble
parameter. It is difficult to evaluate probability of such a state but it may be 
unnecessary. If it existed (with whatever small probability), our large universe 
would evolve out of this microscopically small piece.

If spatial derivatives can
be neglected, field $\phi$ satisfies the following equation of motion:
\be
\ddot \phi + 3H \dot\phi +U'(\phi) = 0
\label{ddot-phi}
\ee
The second term takes into account cosmological expansion and is similar to the liquid 
friction in Newtonian mechanics.
If $H$ is in some sense large (we will specify below the proper conditions), 
then the equation can be reduced to the first order one:
\be
\dot\phi = -U'/3H.
\label{dot-phi}
\ee
Intuitively it is clear that in the case of large friction velocity is proportional
to the force. This is the so called slow roll approximation which works pretty well
in many inflationary scenarios.

If the cosmological energy density, ${ \rho}$, is dominated by slow varying inflaton field
${\phi}$, then the Hubble parameter is equal to
\be 
H^2 = \frac{8\pi U}{3 m_{Pl}^2}
\label{H2-of-phi}
\ee 
From this expression we can estimate the number of e-foldings while $\phi$ 
``lives'' high in the potential $U(\phi)$ and the expansion is approximately
exponential:
\be
N = \int H dt = \frac{8\pi}{m_{Pl}^2}\int\frac{d\phi U(\phi)}{U'(\phi)}\,.
\label{N-efold}
\ee
For the power law potential, ${ U(\phi) = g \phi^n}$, we find:
\be 
N = \frac{4\pi}{n\,m_{Pl}^2}\,(\phi_{in}^2 - \phi^2_{fin}
\approx \frac{4\pi}{n\,m_{Pl}^2}\,\phi_{in}^2.
\label{N-power}
\ee 
For successful inflation $N> 65-70$ is necessary. It implies 
$\phi_{in} > 2.5 n^{1/2} m_{Pl}$. At first sight it looks disturbing if something is
larger than $m_{pl}$. However this is not the case to worry about, because the observable
quantity is the energy density of $\phi$ and it would remain much smaller than $m_{Pl}^4$
because of small mass of $m_\phi$ and the self-coupling constants, as we see in what 
follows. 

For the validity of the slow roll approximation the following two conditions are to be
fulfilled:
\be
\ddot \phi \ll 3 H\dot \phi.
\label{ddot-phi-H-dot-phi}
\ee 
and
\be
\dot\phi^2 \ll 2 U(\phi)
\label{dot-phi2}
\ee 
These conditions are satisfied if
\be 
\mid\frac{ U"}{ U} \mid \ll \frac{8\pi}{3 m_{Pl}^2} 
\label{U-2prime}
\ee 
E.g. for massive free field ${U = m^2\phi^2/2}$ (harmonic potential)
the slow roll approximation would be 
valid if:
\be 
\phi^2 > (4\pi/3)\, m^2_{Pl}
\label{phi2}
\ee   
With $\bm \phi$ exactly at the lower limit the number of e-foldings is not enough but
a slightly larger ${\phi}$ would do the job.
The harmonic potential would not exceed the Planck value if
\be 
\phi^2 < m_{Pl}^4/m_\phi^2
\label{phi2-max}
\ee 
If we take $\phi$ equal to the upper bound, ${\phi_{in} = m_{Pl}^2/m_\phi}$, and 
${m_\phi\sim 10^{-6} m_{Pl}}$, which is demanded by the condition of sufficiently
small density perturbations, the number of e-folding would be huge: 
$N = 10^{13}$. The characteristic time when all this happened is tiny,
${t_{inf} \sim 10^{-31}}$ s.

{\it Problem 19.} Find $N$ and ${t_{inf}}$ for ${U=\lambda \phi^4}$, assuming
$\lambda = 10^{-12}$ and initial $U(\phi) = m_{Pl}^4$.

\subsection{Particle production by inflaton \label{{ss-part-prod}} }
 
During inflation the curvature of the inflaton potential should be smaller than the
Hubble parameter, $U''(\phi) < H_I$, see the slow roll conditions above.
In this regime inflaton monotonically but slowly
moved down to the minimum of $U(\phi)$. Simultaneously decreases the Hubble parameter.
At some moment the second derivative in eq. (\ref{ddot-phi}) became non-negligible
and $\phi$ started to oscillate near minimum with frequency $\omega > H$. The expansion
regime changed from the exponential to matter dominated one, if 
$U(\phi) = m^2_\phi\phi^2$/2,
or to relativistic regime, if $U(\phi) = \lambda \phi^4/4 $.

{\it Problem 20.} Solve eq. (\ref{ddot-phi}) with potential $U= m^2\phi^2$ and find 
cosmological scale factor $a(t)$ and $H(t)$ assuming that the cosmological energy 
is dominated by $\phi$. Use expression (\ref{T-mu-nu-phi}) for the energy-momentum
tensor of $\phi$.

As is well known, a time varying field produces particles to which it is coupled
and especially efficiently those whose mass is smaller than the frequency of the
oscillations. It is exactly what happened with the inflaton. Empty, cold 
universe filled with oscillating $\phi$ exploded with creation of hot relativistic 
particles. It is almost as is described in the Bible: ``Let there be light''.
This moment is proper to call big bang, though it may be not generally accepted 
terminology.

Particle production by inflaton have been first calculated 
perturbatively in ref.~\cite{pert-prod}. It was shown that due to weakness
of inflaton coupling to matter, perturbative  
particle creation is rather slow and the temperature of the universe heating
after inflation is relatively small, much smaller than $\rho_{fin}^{1/4}$,
where $\rho_{fin}$ is the inflaton energy density to the end of inflation.
Let us consider an example when the inflaton is coupled to fermions through
the Yukawa coupling:
\be
L_{int} = g \phi (t) \bar \psi \psi
\label{phi-psi}
\ee 
where $\phi(t)$ is supposed to be classical field satisfying eq.~(\ref{ddot-phi}):
\be
\phi(t) =\frac{m_{Pl}}{\sqrt {3\pi} m_\phi } {\sin m_\phi 
(t+t_0) \over t+t_0} 
\label{phi-of-t}
\ee
In the lowest order of perturbation theory the  amplitude  of 
production of pair of fermions with momenta $k_1$ and $k_2$ is 
\be
A(k_1,k_2) =g\int \,d^4x \sigma (t) \langle k_1,k_2  |\bar  \psi 
\psi |0\rangle =
\frac{(2\pi )^3 g}{\sqrt {4E_1E_2}} \delta ({\bf k_1} +{\bf k_2} ) \bar  u(k_1)  v(k_2) 
\tilde \phi (E_1 +E_2 ),
\label{A-k1-k2}
\ee
where the standard decomposition of $\psi$ and $\bar \psi$ in   
creation-annihilation operators is used, $\bar u$ and $v$ are  the 
Dirac spinors, $E_i =|k_i|$ is the particle energy, and
\be
\tilde \phi (\omega )=\int \,dt e^{i\omega t} \phi (t).
\label{phi-of-omega}
\ee
The probability of particle production per unit volume is 
\be
N_f \equiv {W \over V} = {1 \over V} \int {\,d^3k_1 \,d^3k_2 \over 
(2\pi )^6 } |A|^2 = {g^2 \over \pi^2}  \int_{E>0}  \,dE  E^2  
\tilde \phi (2E)|^2. 
\label{N-f}
\ee
The volume factor  $V$   as   usually   comes   from   the   square   of 
$\delta$-function:
\be
[\delta ({\bf k_1} +{\bf k_2} )]^2 =\delta ({\bf k_1} +{\bf k_2} 
)V /(2\pi )^3
\label{delta-2}
\ee

If $\omega \gg t^{-1}$ then the integration over time in the interval
$\Delta t \gg  \omega ^{-1}$ in eq. (\ref{phi-of-omega}) gives 
$\delta (\omega - 2E)$ and the square of it would be 
$\delta t\,\delta(\omega - 2E)/(2\pi)$. So
we obtain for the rate of the fermion production per unit time and unit volume:
\be
\dot N_f ={N_f \over \Delta t} ={m_1^2  \omega^2  \over  8\pi  } 
={g^2 m_{Pl}^2 \over 24\pi^2 (t+t_0)^2} 
\label{dotN-f}
\ee
This corresponds to the following decay rate of the field $\phi$:
\be
\Gamma _\phi =\frac{\dot N_f}{ N_\phi } 
= \frac{g^2 }{ 4\pi  }  m_\phi\,.
\label{Gamma-phi}
\ee
This is, as one can expect, the decay width of the $\phi$-meson.

Because of the particle production $\phi (t)$  should  decrease 
faster than it given by  eq.~(\ref{phi-of-t})   For  
$\Gamma_\phi  \ll \omega $ this can be taken into account by the substitution
$\phi (t) \rightarrow \phi (t)\,\exp (-\Gamma_\phi t)$.

The rate of thermalisation of the produced fermions is as  a  rule 
larger than the expansion rate. In this case  the  temperature  of 
the plasma can be evaluated as follows. Assume that the  particles 
are produced instantly at the moment when  the  Hubble  parameter 
$H=1/2t$ becomes equal to the decay rate $\Gamma_\phi$ and  that  
$t\gg t_0$. The energy density of the produced fermions is 
\be
\rho _f = {3\Gamma_\phi^2 m_{Pl}^2 \over 8\pi  }  ={3g^4  \over  128 
\pi^3  }  m_{Pl}^2  m_\phi  ^2  
\label{rho-f}
\ee
and   correspondingly   the temperature of the universe heating is 
\be
T_h =\left( \frac{30\rho_f} {\pi ^2  g_*  }  \right)^{1/4}  =  \left( 
{90\over 128\pi^5 g_* } \right)^{1/4} g \sqrt {m_{Pl} m_\phi}. 
\label{T-h}
\ee

For a more accurate evaluation of $T_h$ let us take  into  account 
non instant character of the particle production and the  decrease 
of the amplitude of the oscillations $\phi_0 (t)$ not only because 
of  the  Universe  expansion  but  also  due  to   the   particle 
production. With these  factors  taken  into  account  the  energy 
density of the produced fermions satisfies the equation
\be
\dot  \rho  _f  =  \Gamma_\phi  \rho_\phi  -4H\rho_f,  
\label{dot-rho-f}
\ee
 where  $\Gamma_\phi$  is  given  by  eq. (\ref{Gamma-phi})   and 
$\rho_\phi = m_\phi ^2 \phi_0^2(t) $ is  the  energy  density  of 
oscillating $\phi(t)$. We assume also  that  the  total  energy 
density is equal to the critical one:
\be
\rho_\phi +\rho_f =\frac{3}{8\pi}m_{Pl}^2H^2 = 
\frac{m_{Pl}^2}{6\pi (t+t_0)^2} 
\label{rho-phi-f}
\ee
In the last equation it is assumed also that  the  non-relativistic 
expansion  law  is  valid,  that  is  $\rho_\phi  >\rho_f$.  Thus 
eq. (\ref{dot-rho-f}) takes the form:
\be
\dot \rho_f =\Gamma _\phi \rho_c -(4H +\Gamma _\phi  )\rho_f\,. 
\label{dot-rho-f2}
\ee
Integrating it with the initial condition $\rho_f (0) =0$  we obtain 
for the early MD-stage:
\be
\rho_f  ={\Gamma^2  m_{Pl}^2  e^{-\Gamma  t}  \over 6\pi  [\Gamma 
(t+t_0)]^{8/3}} \int _0^{\Gamma t} \,dx e^x x^{2/3} 
\label{rho-f2}
\ee
This equation is valid till the energy density of the produced fermions becomes 
non-negligible, e.g. till $\rho_f \approx \rho  _\phi  \approx \rho_c /2$. 
This is realised at $\Gamma t  =1.3$  At  that  moment 
$\rho_f \approx g^4 m_\phi^2  m_{Pl}^2  /500  \pi^3$  and  the 
temperature is
\be
T_h \approx \left( {3\over 50\pi^5  g_*  }\right)^{1/4}  g  \sqrt 
{m_\phi m_{Pl}} 
\label{T-h2}
\ee
This is about twice smaller than result (\ref{T-h}).

In many interesting cases perturbative approximation to particle production do not
adequately describe physics of the process. Non-perturbative calculations have been
pioneered in ref.~\cite{non-pert} and further developed in ref.~\cite{kls} with an 
emphasis on the parametric resonance enhacement. Due to that the inflaton decay can
proceed much faster than expected from perturbation theory and the initial particle
state might be far from thermal equilibrium. In particular, heavy particles with
mass larger than would-be temperature could be produced. It was also 
noticed~\cite{grav-part-prod} that production of heavy particles by gravitational
field at the final stage of inflation might be essential as well.

We conclude here that the heating temperature after inflation
is model dependent and most probably is not large,
${ T_{h}< E_{GUT}\sim 10^{15}}$ GeV, {i.e. GUT
era probably never existed.} It solves the problem of overabundant 
magnetic monopoles. Initial hot universe might be far from
thermal equilibrium and very heavy particles could be produced
by the cosmological gravitational field~\cite{grav-part-prod}.

\subsection{Inflationary density perturbations 
\label{ss-perturb}}

The  increase  of  quantum  fluctuations  of  scalar  field  in   the 
exponentially  expanding   space-time   gives   rise   to   density 
inhomogeneities in the Universe. Physically  this  phenomenon  is 
connected with different  moments  of  the  end  of  inflation  in 
different space points (in an appropriate reference  frame)
due to small spatial fluctuations of the inflaton. 
Below we will follow the simple presentation of review~\cite{blau-guth}.
For detailed rigourous treatment see book~\cite{mukhanov}

Exponential  expansion  transforms  microscopically   small   wave 
lengths of quantum fluctuations into  astronomically  large  ones 
and  so   produce   natural   candidates   for   initial   density 
inhomogeneities which could  be  the  seeds  of  the  large  scale 
structure formation. In a sense this task is even  over-fulfilled  because 
the inhomogeneities $\delta \rho$ prove to be too  large  for
natural values of the parameters of the theory. In  particular  to 
get  $\delta \rho /\rho \approx 10^{-4}$  on  the 
galactic scale the self-coupling of the inflaton should be 
$\lambda < 10^{-12}$.

Since we expect that the density fluctuations have originated from quantum
fluctuation during inflation, we need to say a few words about scalar field
quantisation in De Sitter space-time. We assume that the field is massless
because we are interested in $m_\phi \ll H$.

Field $\phi$ is quantised in the De Sitter  space-time  along  the  same 
lines as in flat space-time. The starting point is  the  expansion 
in creation-annihilation operators:
\be
\phi (x,t) =\int {\,d^3k \over (2\pi  )^{3/2}  \sqrt  {2\omega_k 
}}[a_k e^{-ikx}\phi (t) + h.c. ],
\label{phi-expand}
\ee
where $\omega _k =\sqrt {k^2 + m^2}$ and $\phi _k  (t)$  satisfies the Furrier
transformed free equation of motion (in curved space-time):
\be
\ddot \phi_k +3H\dot \phi_k + \frac{k^2}{ a^2} \phi_k =0.
\label{ddot-phi-of-k}
\ee
Operators $a_k$ and $a_{k'}^+$ obey the standard 
commutation relation 
\be
[a_k,a_{k'}^+ ]=\delta^{(3)} (k-k').
\label{a-commute}
\ee
For massless scalar field the solution of eq. (\ref{ddot-phi-of-k})  is  expressed 
through the Bessel functions:
\be
\phi_k (t)=C_{1k} (k\tau  )^{3/2}  [J_{-3/2}  (k\tau  )  +C_{2k} 
J_{3/2} (k\tau )],
\label{phi-k-sol}
\ee
where  $\tau  = \exp(-Ht)/H$ is the conformal time and the coefficients  $C_{ik}$ are 
determined  by  the  matching  $\phi_k$  to  the  corresponding 
expression in   flat space-time,
$\phi_k \rightarrow \exp(-i\omega_kt)$  for  $H\rightarrow  0$  and 
$\tau \rightarrow H^{-1} -t$. Substituting asymptotic  values
of the Bessel functions for $k\tau \rightarrow  \infty$  into  eq. (\ref{phi-k-sol}) 
we find
\be
C_{1k} &=& \sqrt {{\pi \over 2}}{H\over k}\, e^{i\alpha_k} \nonumber\\
C_{2k} &=&-i,
\label{C1k-C2k}
\ee
where $\alpha$ is a constant phase.

In the limit of large $t$ (or $\tau  \rightarrow  0$)  one  easily finds:
\be
| \phi_k |^2 \approx {H^2 \over k^2} \left( 1+ {k^2 \over  H^2} 
e^{-2Ht} \right)
\label{phi2-of-k}
\ee

Let us extract from the  inflaton  field  $\phi  (x,t)$  classical 
homogeneous part $\phi_0(t)$:
\be
\phi (x,t) = \phi_0 (t) + \delta \phi (x,t), 
\label{phi-of-x-t}
\ee
where $\delta \phi (x,t)$ describes small quantum fluctuations.  On 
inflationary stage it satisfies the equation:
\be
\delta \ddot \phi +3H\delta  \dot  \phi  -e^{-2Ht}  \partial_i^2 
\delta \phi -{\partial^2 V(\phi_0 )\over \partial  \phi^2}  \delta 
\phi =0.
\label{ddot-delta-phi}
\ee
For large $Ht$ the third term in the equation can be neglected and 
$\delta \phi$ satisfies the same equation as $\dot  \phi_0$  does,
see eq. (\ref{ddot-phi}). 
Equation (\ref{ddot-delta-phi})
has two solutions. One of them  decreases  as  $\exp (-3Ht)$ for 
$\partial^2 V /\partial \phi^2 \ll  H^2$.  The  second 
solution varies relatively slowly. So at  large  $t$  the  first 
solution can be neglected and we can write 
\be
\delta \phi (x,t) =-\delta \tau  (x)  \dot  \phi_0  (t).  
\label{delta-pho-of-x-t}
\ee 
If $\phi$ is small this is equivalent to $x$-dependent retardation 
of the classical field motion to the equilibrium point:
\be
\phi (x,t) =\phi_0 \left( t-\delta \tau (x)\right).
\label{phi-retard}
\ee
Correspondingly  inflation  ends  at  different 
moments in different space points. This is the physical reason for 
generation of density perturbations. Since the energy  density  in 
the universe during inflation is  dominated  by  the  inflaton 
field $\phi$, one can write  $\rho  (x,t)  =  \rho  \left(t-\delta  \tau (x)\right) $ 
forgetting possible  subtleties  connected  with  the 
freedom in coordinate choice. Thus we get
\be
{\delta \rho \over \rho }=-\delta \tau {\dot \rho \over  \rho  } 
=4H\delta \tau (x). 
\label{delta-rho-over-rho}
\ee
At the last step the relation $\dot  \rho  /\rho  =-4H$  has  been 
used. It is valid on RD-stage which by assumption  was  formed  in the
heated universe  when inflation was over. 
It is convenient to make the Furrier transform with a different integration
measure:
\be 
\langle \delta \phi (x,t)^2 \rangle =\int {\,d^3k \over  k^3}  | 
\Delta \phi (k,t)|^2. 
\label{delta-phi-2}
\ee
The density fluctuations can be now expressed in terms of 
$\Delta \phi_k$ as
\be
{\delta \rho \over \rho} =4H {\Delta \phi_k \over \phi_0}.  
\label{delta-rho-2}
\ee
The Fourier amplitude of the fluctuations is evaluated with the help 
of eqs.~(\ref{phi-k-sol},\ref{phi2-of-k}) as:
\be
\Delta \phi (k,t) ={H\over 4\pi^{3/2} }\left( 1+ {k^2 \over H^2} 
e^{-2Ht} \right)^{1/2}. 
\label{delta-phi-of-r-t}
\ee

Recall that this expression is  valid  for  massless  free  field. 
Since $\delta \phi (x,t)$ satisfies eq. (\ref{ddot-delta-phi}) the 
approximation is valid till 
$k^2 \exp (-2Ht) \geq |\partial^2 U /\partial \phi^2 |$.
On the other hand, the proportionality condition (\ref{delta-pho-of-x-t})  
is  valid 
in the opposite limit when one can neglect the decreasing part  of 
the solution. So for the evaluation of  the  density  fluctuations 
one has to substitute expression (\ref{delta-phi-of-r-t}) 
in the boundary region:
\be
t=t_b(k) =\frac{1}{2H} \ln  \frac{k^2}{|\partial^2  U  /\partial 
\phi^2 |}
\label{t-b}
\ee
One should remember that $U(\phi)$  depends  on $t$  through  
$\phi =\phi_0 (t)$ so eq. (\ref{t-b}) implicitly determines $t_b$.
Finely we find:
\be
\frac{\Delta \rho}{\rho }  \mid_h  ={H^2  \over  \pi^{3/2}  \dot 
\phi_0  \left[  t_b  (k)  \right)  }  \left[  1+   \frac{1}{ H^2} 
\frac{\partial^2 V \left( \phi_0 (t_b (k)) \right)}{ \partial \phi^2 
} \right]^{1/2}. 
\label{delta-rho-3}
\ee
This expression for the fluctuation spectrum  is  valid  till  the 
fluctuation wave length reaches the horizon. This is indicated  by 
sub-h in the left hand side. 

Since $\phi  (t)$  is  a  slowly  varying  function  of  time,  the 
fluctuation spectrum weakly depends  on  $k$.  This  flat 
spectrum is known as the Harrison-Zel'dovich  spectrum  
For a  satisfactory  description  of  the 
universe structure the fluctuations should have the value  $\Delta 
\rho /\rho |_h \approx 10^{-4}$.

As an example let us consider density  perturbations  in  the 
model with the potential $U(\phi )  =-\lambda  \phi^4  /4$.  
This potential is not bounded from below, so we can trust it only
for sufficiently low $\phi$. 
Homogeneous classical  field  $\phi_0 (t)$ satisfies the equation:
\be
\ddot \phi +3H\dot \phi -\lambda \phi^3 =0. 
\label{ddot-phi-lambda4}
\ee
We assume that the term $\ddot \phi$ can be  neglected,
see subsection~\ref{ss-models},  and  check 
the validity of this approximation on the explicit solution.  
The latter has the form:
\be
\phi  (t)  =\left(  {3H  \over  2\lambda}   \right)^{1/2}   (t_f 
-t)^{-1/2} \,,
\label{phi-of-t-lambda}
\ee
where $t_f$ is approximately the moment when inflation ended. In  a 
realistic model $U(\phi)$ is bounded from below, so $\phi (t)$ does 
not tend to infinity. Still solution (\ref{phi-of-t-lambda}) gives  a  satisfactory 
approximation up to $t$  slightly  less  than  $t_f$.  After  that 
moment  the  rise  of  $\phi$  should turn into oscillations around 
the equilibrium point.

The neglect of $\ddot \phi$ is justified if the condition
\be 
{\ddot \phi \over 3H\dot \phi }={1 \over 2H(t_f -t)} \ll 1
\label{ddot-phi-over-3H}
\ee 
is fulfilled. Substituting solution (\ref{phi-of-t-lambda}) into 
eq. (\ref{delta-rho-3}) we arrive to 
\be
\frac{\Delta  \rho}{ \ rho}   \mid_h   =\left(  \frac{8}{ 3\pi^3 
}\right)^{1/2} \lambda^{1/2} [H(t_f-t_b)]^{3/2} ,
\label{delta-rho-4}
\ee
where in accordance with eq. (\ref{t-b}):
\be 
Ht_b =\ln \frac{k}{H} +\frac{1}{  2}  \ln \frac{2H(t_f  -t_b)}{ 9}.  
\label{Ht-b}
\ee
The second term in this expression  is  evidently  small. 
$Ht_b$ can be approximately expressed through  the  comoving  wave 
vector $k$  and  the  corresponding  to  it  present-day  physical 
scale $l_0$ as:
\be 
l_0 = \frac{2\pi a(t_0)}{ka(t_f)  }e^{Ht_f}  \approx  
\frac{2\pi  T_h }{ kT_0 }e^{Ht_f},
\label{l-0}
\ee
where $a(t)$ is the scale factor, $T_0$ is the  present-day  value 
of  the CMB temperature,  and  $T_h$  is   the 
heating temperature at the end of inflation.  It  is  taken  into 
account that     the     physical      momentum changes 
inversely proportionally  to  the   scale   factor.   In particular   during 
inflationary stage $p=k\exp(-Ht)$  and  during  Friedman  stage  $p$ basically
decreases as inverse temperature. 

As a result 
\be
Ht_f =\ln \frac{kl_0T_0 }{ 2\pi T_R }
\label{H-tf}
\ee
and finally we obtain
\be
\frac{\Delta \rho }{ \rho }\mid_h = \left( \frac{8}{ 3\pi^3} \right) 
\lambda^{1/2} \ln^{3/2} \frac{l_0 }{ b},
\label{delta-rho-fin}
\ee
where $b=(2\pi T_h /HT_0)$. For $T_h =10^{15}$ GeV which is possibly rather high, 
but not unreasonable, we  obtain  $b\approx  15\,{\rm }m  \approx 
1.5\cdot 10^{-15}$ years. Hence the density fluctuations on the 
galactic scale $l_0 =10^6$ years are 
\be
\Delta \rho /\rho \approx 10^2 \lambda^{1/2}
\label{delta-rho-gal}
\ee
So that $\lambda $ should be tiny, $\lambda  \approx  10^{-12}$,  to 
give rise to a proper value  of  the  fluctuations,  
$\Delta  \rho /\rho  \approx  10^{-4}$.  
This  is  a   common   shortcoming   of 
inflationary models. Such a small value means in  particular  that 
the inflaton should be a gauge singlet or more complicated scenarios
are necessary. Otherwise  the  interactions 
with gauge bosons would generate too big effective coupling, 
$\lambda_{eff} \phi^4$ with $\lambda_{eff} \approx \alpha^2 \approx 10^{-4}$.

The perturbation spectrum is not exactly flat but slightly deviates from the 
Harrison-Zeldovich one. This is a general feature of all inflationary
models.

Generation of gravitational waves during inflation can be considered along the
same lines. Moreover, the equation of motion of spin eigenstates of gravitons 
coincides with the equation of motion of massless scalar field. It may be
instructive to note that gravitational waves are not created in eternal 
De Sitter background. One can check that the Bogolyubov coefficients which
describe particle production are trivial. But when expansion is changed
from exponential to, say, power law, the production of particles and waves
becomes possible.

\subsection{Inflationary conclusion \label{ss-concl}}

Inflation seems to be practically an experimental fact. It nicely explains the
observed features of the universe. In particular, the observed spatial flatness
today, $\Omega = 1\pm (\sim 10^{-2})$ 
and very exact fine-tuning at, say, BBN, $\Omega = 1\pm 10^{-15}$,
naturally arises because of exponential 
expansion in the early universe. The observed spectrum of density fluctuations
is predicted to be nearly flat, but not exactly flat. This is another successful
prediction of inflation. The only ``missing link'' is not yet observed, namely, long 
gravitational waves, which may be accessible to LISA. However, one 
should keep in mind that though an observation of such
waves will be a strong argument in favour of inflation but if they are not observed,
inflation still will not be killed, because the amplitude of inflationary
gravitational waves is model dependent and may be so small that they will escape 
observation by the near future antennas.

\section{Baryogenesis \label{s-bg}}

The {observed} part of the universe is 100\% C(CP)-asymmetric. 
Up to now no astronomically significant objects consisting antimatter 
have been detected.
There is only matter and {no antimatter}, 
except for a small number of antiprotons and 
positrons most probably of {secondary} origin.
From the bounds of the flux of 100 MeV gamma rays one can conclude that
the nearest {galaxy}, if dominated by antimatter,
should be at least at $\sim$10 Mpc~\cite{steigman-76}.
However we cannot say much about galaxies outside of our super-cluster.
Observed colliding galaxies at any distance or galaxies in 
the common cloud of intergalactic gas are of 
{the same kind of matter (or antimatter?).} In particular, the fraction of
antimatter in two colliding galaxies in Bullet Cluster is bounded  
by $n_{\bar B}/n_B < 3\times 10^{-6}$~\cite{steigman-08}.
In charge symmetric universe the nearest antimatter domain should be practically 
at the cosmological horizon,
{${ l_B >}$ Gpc}, because of {very efficient annihilation} at an
early stage~\cite{CdRG}. {Still smaller clumps of antimatter are allowed 
in our neighbourhood.}

However, one should bear in mind that these bounds are true if antimatter makes  
exactly the same type objects as the {\it observed} matter.
For example, compact objects made of antimatter may escape 
observations and be quite abundant and almost at hand.

A natural question arises in this connection: is matter predominance
accidental (a result of asymmetric initial conditions) or dynamical?
Inflation proves that it is dynamical, originated from some physical processes
with non-conserved baryon number.
Sufficient inflation should last at least {($\sim$ 70 Hubble times)} with
practically constant Hubble parameter. This could happen only if the energy 
density was approximately {constant}, see eq.~(\ref{H2}).
However, if baryons are conserved, the energy density associated 
with baryonic number}) cannot be
constant and inflation could last at most {4-5 Hubble times.}
Indeed if baryonic charge were conserved then it would remain constant in the
comoving volume, i.e. $B\sim 1/a^3$. The energy density of bearers of this 
baryonic number cannot stay constant as well but should evolve as 
$\rho_B \sim 1/a^n$, where $n=3$ for non-relativistic matter and $n=4$ for
relativistic matter.

Using the observed value of the cosmological density of the baryons
and assuming its conservation, we find that at the early relativistic epoch (RD),
e.g. at BBN the energy density related to this baryonic number should be
${ \rho_B \approx 10^{-7}\rho_{tot} }$. During evolution at RD-stage the ratio
$\rho_B/\rho_{tot}$  remained practically constant. Let us go backward in time
till inflation. At inflation $ \rho_{tot} \approx const$, but baryons excluded.
The energy density of the latter, if their number is conserved, should evolve as
$ \rho_B \sim \exp (-4Ht)$. It means that 4-5 Hubble times back
sub-dominant baryons were dominant and
{${ \rho_{tot} \approx \rho_B}$}, which could not stay constant more that 4-5
Hubble times. This is surely insufficient for solution of cosmological problems
discussed in the previous section.

It was suggested by Sakharov in 1967~\cite{sakharov} 
that the cosmological baryon asymmetry
could be generated dynamically in the early universe if the following three
conditions are fulfilled: \\
1. {Non-conservation of baryons}\\
2. {Breaking of C andCP symmetries.}\\
3. {Deviation from thermal equilibrium.}\\
Note in passing that none of them is obligatory -- we present examples below. 
But first let us discuss normal baryogenesis and validity of these three conditions.

I. Non-conservation of baryons is justified theoretically. Grand unified theories,
non-minimal SUSY, and electoweak (EW) theory predict that baryonic number is not
conserved, $\Delta B \neq 0$. However, at the present time this prediction is not
confirmed by direct experiment. Despite an extensive search, only upper bounds on
proton life-time and period of neutron-antineutron oscillations are established.
The only ``experimental piece of data'' in favour of baryon non-conservation
is our universe: {we exist, ergo baryons are not conserved.} Half of century ago
from the same experimental fact, our existence, an opposite conclusion of baryon
conservation was deduced. Theory is an important input in understanding of what we
observe.

II. C and CP violation are discovered and confirmed in direct experiments.
At the first part of the XXth century the common belief was
that physics was invariant with respect to separate
action of all three transformations: mirror reflection, P, charge conjugation, C, and
time reversal, T. The weakest link in this chain of discrete symmetries was P,
found to be broken in 1956~\cite{p-viol}. 

It was immediately assumed that the world was symmetric with respect to the
combined transformation from particles to mirror reflected antiparticles, CP.
Both P and C are 100\% broken in weak interactions but still some symmetry 
between particles and antiparticles was saved. 
This symmetry crashed down pretty soon, in 1964~\cite{cp-viol}.
After this discovery life in the universe became possible. 

Why CP-breaking is necessary for generation of cosmological baryon asymmetry
but C-breaking is not enough? A formal answer to this question is the following.
Let us assume first that C is conserved and that the universe was
initially in $C$ eigenstate, i.e.
\be
C|u\rangle = \eta |u\rangle
\label{C-u}
\ee
where $|u\rangle$ is the wave function of the universe and
${|\eta|=1}$ is a constant.
This means, in particular, that the universe had 
initially all zero charges because
charge operator anticommute with C-transformation.
{May some non-zero charge, e.g. ${B}$, be generated dynamically?} 
The answer is negative because due to C-invariance the Hamiltonian of the system
commutes with C-operator, $\left[ C, {\cal H}\right] = 0$. 
The time evolution of ${B}$ is given by:
\be
B(t) = \langle u | e^{-i {\cal H}t} J^B_0 e^{i{\cal H}t} |u\rangle.
\label{B-of-t}
\ee
Let us insert  into this equation the unity operator ${ I = C^{-1} C}$:
\be
B(t) =\langle u| I 
 e^{-i{\cal H}t} I J^B_0 I e^{i{\cal H} t} I
| u \rangle = - B(t), 
\label{C-B-of-t}
\ee
taken that ${{ C J^B_0 C^{-1} = -J_0^B}}$.  Thus in C-conserving 
theory ${B(t)=B_{in}=0}.$

The same arguments with ${ CP}$ instead of ${C}$ prove
that charge asymmetry cannot be 
generated, if CP is conserved and
{the universe is an eigenstate of CP:} 
\be
CP |u\rangle = \eta |u\rangle.
\label{CP-u}
\ee
In rotating universe charge asymmetry might be
generated even if CP is conserved. { Global rotation can be 
transformed into baryonic charge}! However, it seems difficult to realise
such an idea. {New long-range interactions, possibly very unusual,
are needed.

For B-generation in elementary (local) processes no assumption about the universe 
state is necessary:
{if CP is conserved, no asymmetry is generated through particle 
decays or reactions.} We will discuss this below in concrete examples.

At the present time only CPT-symmetry survived destruction. It is the only one
which has rigorous theoretical justification, {CPT-theorem,}
based on solid ground:  of Lorenz-invariance, canonical spin-statistics relation,
and positive definite energy.
Still models {without CPT} are considered, e.g. 
for explanation of some neutrino anomalies and for baryogenesis.

III. Thermal equilibrium is always broken for massive particles, but usually very
little. To estimate the effect let us approximate
the collision integral in kinetic equation (\ref{kin-eq}) as 
$I_{coll} =\Gamma (f_{eq} - f)$, where $\Gamma$ is the interaction rate.
Let us assume that $\Gamma$ is large so that the deviation from equilibrium is
small, $\delta f/f_{eq} \ll 1$. Substituting into the l.h.s. of kinetic equation
$f_{eq}$, as is done in eq. (\ref{lhsm-zero}), we find:
\be 
\frac{\delta f }{f_{eq}} \approx \frac{H m^2}{\Gamma T E} \sim 
\frac{T m^2}{ \Gamma E {m_{Pl}}}.
\label{delta-f-over-f}
\ee 
Since the Planck mass is very large the deviation from equilibrium might be
significant only at large temperatures or tiny $\Gamma$. However, if the fundamental
gravity scale is near TeV~\cite{add}, equilibrium could be strongly broken even
at the electroweak scale.

Another source of deviation from equilibrium in the cosmological plasma
could be first order phase transition from, say, unbroken to broken symmetry phase
in non-abelian gauge theories with spontaneous symmetry violation. There might be
a rather long non-equilibrium period of coexisting two phases.

There are plenty scenarios of baryogenesis each of them
one way or other performing a rather modest task  
explaining only one number, the observed asymmetry:
\be
\bm{\beta} = (n_B-n_{\bar B}) /n_\gamma = 6\cdot 10^{-10},
\label{beta}
\ee
found from the analysis of two independent measurements: of light element 
abundances created at BBN and of angular fluctuations of CMB.

It is a great challenge to astronomers to check if 
${\beta}$ is constant or it may vary at different space points,
{${ \beta = \beta (x)}$}.
What is characteristic scale 
${{\ l_B}}$ of variation of baryonic number density?
May there be astronomically {large domains of antimatter} nearby or only
very far away? Answers to these questions depend, in particular, 
upon mechanism of CP violation realized in cosmology, which
are described below. For more detail see lectures~\cite{ad-cp}. 
There are three possibilities for CP-breaking in cosmology:

1. {\it Explicit}, realized by complex coupling constants in
Lagrangian, in particular, complex 
Yukawa couplings transformed by the vacuum expectation value of the
Higgs field ${ \langle \phi \rangle \neq 0}$ into  
non-vanishing phase in CKM-mixing matrix. However, in the minimal
standard model (MSM)  based on $SU(3)\times SU(2)\times U(1)$
CP-violation at $T\sim$ TeV is too weak, at least by 10 orders of magnitude,
to allow for generation of the observed baryon asymmetry.
Indeed, CP-violation in MSM is absent for two quark families because the phase
in quark mass matrix can be rotated away. So at least 
three families are necessary. It could be an 
anthropic explanation why we need three generations.

If {masses} of different up or down quarks {are equal,} 
CP violation can be also {rotated away} because the unit matrix is invariant
with respect to unitary transformations.
If  {the mass matrix is diagonal} in the same representation as flavour 
matrix, CP-violation can also be {rotated away}.
Thus CP-breaking is proportional to the product of 
{the mixing angles} and to 
{the mass differences} of all down and all up quarks:
\be
A_- \sim \sin \theta_{12}\, \sin \theta_{23} \sin \theta_{31}\, \sin\delta\,
(m_t^2-m_u^2)(m_t^2-m_c^2)(m_c^2-m_u^2)  \\
\nonumber
(m_b^2-m_s^2)(m_b^2-m_d^2)(m_s^2-m_d^2)/{{\ M^{12}}}.
\label{A-}
\ee
At high ${ T\geq }$ TeV, where electroweak baryon nonconservation is operative, 
the characteristic mass ${ M\sim 100}$ GeV and $A_- \sim 10^{-19}$.
So for successful baryogenesis an extension of MSM is necessary.

2. {\it Spontaneous CP violation}~\cite{cp-spont}, which could be realized
by a complex scalar field $\Phi$ with CP-symmetric potential,
with two separated minima at $\langle \Phi \rangle = \pm f $. 
The Lagrangian is supposed to be CP-invariant but these two vacuum states have
the opposite signs of CP-violation. Such CP-breaking is
locally indistinguishable from the explicit one but globally 
leads to charge symmetric universe with equal amount of matter and antimatter.
As we mentioned at the beginning of this section, the antimatter domain should be
very far at {${ l_B \geq} $ Gpc.}  Moreover, there is another problem with this mechanism,
namely domain walls between matter and antimatter domains could destroy the observed
homogeneity and isotropy of the universe~\cite{zko}. To avoid the problem
a mechanism of the wall destruction is necessary.

3 {\it Stochastic or dynamical.} If a
complex scalar field ${ \chi}$  was displaced 
from its equilibrium  point in the potential, e.g.
by quantum fluctuations at inflation, and did not relaxed down to equilibrium
at baryogenesis, it would create CP-violation proportional to the amplitude of the
field but without problems of spontaneous CP-violation. Later, after baryogenesis
was over, $\chi$ would relax down to zero. So domain walls do not appear. 
Inhomogeneous $\beta (x)$ with domains of matter and antimatter can be created
with such CP-violation. Their size depends upon the details of the scenario. 

There is a long but probably incomplete list of different scenarios of 
baryogenesis (BG):\\
{1. Heavy particle decays~\cite{sakharov}.}\\
{2. Electroweak BG~\cite{krs}.}
Too weak in MSM but may work with TeV gravity.\\
3. Baryo-through-leptogenesis~\cite{fy}.\\
4. SUSY condensate BG~\cite{ad}.\\
5. Spontaneous BG~\cite{ck}.\\
6. BG by PBH evaporation~\cite{bh-bg}.\\
7. Space separation of $B$ and ${{\bar B}}$ at astronomically large 
distances~\cite{omnes}, which is probably not effective. However
anti-baryons might be removed  from our into higher dimensions~\cite{dgp}
or predominantly accumulated inside quark nuggets~\cite{e-zh}. \\
{7. BG due to CPT violation~\cite{bg-cpt}.}\\
In all these scenarios {new physics beyond minimal standard model is necessary.}
In what follows we will very briefly describe some of these scenarios. 
More details can be found in the reviews~\cite{bg-reviews}.


{\it Heavy particle decay BG} is naturally realized in grand unification
theories, GUTs, where gauge bosons $X$ with mass around 
$10^{16}-10^{15}$ GeV are present. These
bosons can decay e.g. into, $qq$ and $\bar q l$ pairs where baryon number is 
evidently not conserved. Due to large mass of $X$ the deviation from equilibrium
could be significant, CP-violation might be sufficiently large (we know nothing 
about it) and the mechanism could be efficient enough to generate the observed asymmetry.
The problem with GUTs is that the temperatures of the GUT scale might not be reachable
after inflation. On the other hand, baryogenesis might proceed with under-abundant 
$X$--bosons created out of equilibrium.

Particles and antiparticles can have {different decay rates} into charge 
conjugated channels if C and CP are broken, while
the {total widths are equal due to CPT invariance.}
If only C is broken, but CP is not, then partial widths, summed over
spins, are the same because CP-invariance implies:
\be
\Gamma \left( X\rightarrow f, \sigma \right) =
\left( \bar X\rightarrow \bar f, -\sigma \right). 
\label{Ganna-barGamma}
\ee

{If both C and CP are broken, partial widths may be different,}
but the effect takes place happen in higher orders of perturbation theory.
In lowest order the amplitudes of charged conjugated processes must be equal,
${ A=\bar A^*}$, because of hermicity of Lagrangian.
The same would be also true for higher order contributions if they were 
real. An imaginary part is generated by re-scattering in the final state
(with non-conservation of $B$ or $L$), as can be 
seen from the S-matrix unitarity condition:
\be
i(T_{if}-T_{if}^\dagger ) = 
- \sum_n\, T_{in} T^\dagger_{nf}
= -\sum_n\, T^\dagger_{in} T_{nf}
\label{Im-A}
\ee

Let us consider an example of $X$-boson decays into the channels:
\be
X \rightarrow qq, \,\,\, X \rightarrow  q\bar l,\nonumber\\
\bar X \rightarrow \bar q \bar q, \,\,\,
\bar X \rightarrow \bar q  l \, .
\label{X-decays}
\ee
and assume that the partial widths are different due to C and CP violation:
\be{
\Gamma_{X\rightarrow qq} = (1+\Delta_q) \Gamma_q, \,\,
\Gamma_{X\rightarrow  q \bar l} = (1-\Delta_l) \Gamma_l}, \nonumber \\
\Gamma_{\bar X \rightarrow \bar q \bar q } = (1-\Delta_q) \Gamma_q, \,\,
\Gamma_{\bar X \rightarrow \bar q l} = (1+\Delta_l) \Gamma_l.
\label{Delta-Gamma-X}
\ee
Here $\Gamma \sim \alpha$ and $\Delta \sim \alpha$, where $\alpha \sim 1/50$ is the
fine structure constant at GUT scale. The asymmetry is proportional to
$ \beta \sim (2/3)(2\Delta_q -\Delta_l)$. Its magnitude can be roughly estimated as
\be
\beta \sim \frac{\delta f}{f}\,\frac{\Delta \Gamma}{\Gamma} 
\sim \frac{m}{m_{Pl}}
\label{beta-GUT}
\ee 
Small numerical coefficients omitted here would diminish the result. For example,
the subsequent entropy dilution by about 1/100 is not included.
For successful lepto/baryo-genesis the mass of the decaying particle 
should be {larger than ${10^{10}}$ GeV},
or ${m_{Pl} \ll 10^{19}}$ GeV).

{\it Problem 21.}
How the charge asymmetry generated in heavy particle decay vanishes in equilibrium?
It is stated in the literature that the inverse decay does the job. However one can
see that it is not so because using CPT, one finds:
\be
\Gamma_{\bar q \bar q \rightarrow \bar X } = (1+\Delta_q)\Gamma_q,
\,\,
\Gamma_{\bar q l \rightarrow \bar X } = (1-\Delta_l)\Gamma_l,
\nonumber \\
\Gamma_{ q  q \rightarrow  X } = (1-\Delta_q)\Gamma_q,
\,\,
\Gamma_{ q \bar l \rightarrow  X } = (1+\Delta_l)\Gamma_l.
\label{CPT-decay}
\ee
Thus direct and inverse decays produce the {same} sign of baryon
asymmetry!

{\it Electroweak baryogenesis} is very attractive because 
all the necessary ingredients are present in the minimal standard model.
CP is known to be broken, but, as we have seen above, very weakly.
Baryonic number is non-conserved because of {nonabelian}
chiral anomaly. At zero T baryon nonconservation is exponentially suppressed as
{${ \exp (-2\pi/\alpha)}$}~\cite{t-hooft}, 
because of barrier penetration between 
different vacua. However, it is argued that at high T it is possible to go over 
the barrier, by formation of classical field 
configuration, sphalerons. We do not know how to calculate  
the probability of production of large coherent field
configurations in elementary particle collision  but lattice simulations show that their
production might be efficient and sphalerons could have thermal equilibrium abundance. 
Unfortunately the deviation from equilibrium of massive particles at EW scale is tiny
and the first order electroweak phase transitions seems to be excluded because of
heavy Higgs boson. So electroweak baryogenesis, though very attractive is not
efficient enough,  but it may be operative with TeV gravity.

{\it Baryo-through-leptogenesis} is probably the most popular mechanism today. 
It started from creation of lepton asymmetry by $L$-nonconserving decays of heavy,
${ m\sim 10^{10}}$ GeV, Majorana neutrino, analogously to GUT, and subsequent 
transformation of the lepton asymmetry into baryonic asymmetry by CP symmetric
and $B$ non-conserving and
${(B-L)}$ conserving electroweak processes. The mass matrix of three flavour
light and heavy Majorana neutrinos has 6 independent phases, three in the sector of 
light neutrinos and 3 in heavy ones. They are unknown and allowed to be of order
unity. 

{\it Primordial black hole evaporation} does not demand B-nonconservation at
particle physics level for generation of the baryon asymmetry. Of course,
thermal evaporation cannot create any charge asymmetry.
However the spectrum is not exactly black but is modified due to propagation
of the produced particles in gravitational field of BH.
Moreover, an interaction among the produced particles is essential. 
Let us assume that a meson $A$ is created at the horizon and decays as:
\be
A\rar H+\bar L \,\,\,{\rm { and}}\,\,\, A\rar \bar H + L,
\label{A-decay}
\ee 
where $H$ and $L$ are a heavy and light baryons, e.g. $t$ and $u$ quarks, respectively.
Due to CP-violation the branching ratios of these decays may be different.
{Back-capture of ${ H}$ by gravitational field of the black hole is larger than 
that of ${ L}$.} {Thus some net baryon asymmetry in the external world could be 
created.}
If the cosmological energy density of black holes at production was small,
$\rho_{BH}/\rho_{tot} =\epsilon \ll 1$,  then at red-shift
${ z=1/\epsilon}$, with respect to the production moment,
the non-relativistic BHs would dominate. Their evaporation could provide
the necessary baryon asymmetry and {reheat} the universe.

For a numerical estimate let us present some simple formulae, omitting numerical
factors of order $1-10$.
The black hole temperature is essentially given by the only available
parameter with dimension of length, i.e. by its gravitational radius:
\be
T_{BH}\sim 1/r_g \sim m_{Pl}^2/M_{BH}.
\label{T-BH}
\ee 
The luminosity of the body with temperature $T_{BH}$ and radius $r_g$ is:
\be
L_{BH} \sim T^4 r_g^2 \sim m_{Pl}^4/M^2_{BH}.
\label{L-BH}
\ee 
Correspondingly the BH life-time is equal to:
\be
\tau_{BH} \sim M_{BH}^3/m_{Pl}^4.
\label{tau-BH}
\ee 
For example, if ${M_{BH} = 10^{15}}$g, its life-time is equal to the universe age,  
${ \tau_{BH} \approx t_U \sim 10^{10}}$ years. For our case much lighter BHs are needed.
Let us assume that primordial BHs were formed at 
${ T_{BH} = 10^{14}}$ GeV and their mass was equal to the mass inside the cosmological
horizon at that moment, {${M_{BH} = m_{Pl}^2 t \approx 10^4 }$g.} The life-time
of such BHs would be {${ \tau_{BH} \sim 10^{-16}}$ sec,}
which corresponds to cosmological temperature 
\tcred{${T \sim 10^5}$ GeV} and red-shift from the moment 
when horizon mass was equal to ${M_{BH}}$, was about 
{${10^{9}}$.} In other words, if the mass fraction of BHs at production was 
$10^{-9}$, then at the moment 
of their evaporation they would dominate the cosmological energy density and
could create observed baryon asymmetry even if the fraction of the baryon number density
was small in comparison with the total number density of the evaporated particles.

{\it Spontaneous baryogenesis} may operate in thermal equilibrium. 
Explicit CP-violation is not obligatory.
It is assumed  that a global $U(1)$-symmetry associated with baryonic number
is spontaneously broken. The Higgs-like scalar boson acquires non-zero vacuum expectation
value and its phase becomes massless Goldstone boson, $\phi = \eta\, \exp (i\theta)$.
In the broken phase the Lagrangian can be written as:
\be
{\cal L} =\eta^2 (\partial \theta )^2 + 
\partial_\mu \theta j^B_\mu
- V(\theta) + i\bar Q \gamma_\mu \partial_\mu Q
i\bar L \gamma_\mu \partial_\mu L + (g \eta \bar Q L + h.c.).
\label{L-theta}
\ee 
In the case of homogeneous $\theta (t)$ the second term looks like chemical potential, 
${{ \dot\theta n_N}}$. However, in 
reality it is not true, because chemical potential is introduced into Hamiltonian
but for derivative coupling {${{{\cal L} \neq {\cal H}}}$.}

If the potential ${ V(\theta)}=0$, i.e. in purely Goldstone case, we can 
integrate the equation of motion:
\be
2 \eta^2 \partial^2 \theta = -\partial_\mu j_\mu ^B
\label{d2-thetea}
\ee
and obtain:
\be
\Delta n_B = - \eta^2 \Delta \dot \theta,
\label{Delta0nB}
\ee
i.e. non-zero baryon asymmetry in thermal equilibrium and without
explicit CP-violation. The latter is created by initial 
${{ \dot\theta\neq 0}} $.

In realistic situation ${{ \dot\theta}}$ is small (because inflation kills all
motion) and  the pseudogoldstone case, i.e. non-zero $V(\theta)$
could be more efficient.
Now the equation of motion for $\theta$ takes the form
\be
\eta^2 \ddot \theta +3H\dot \theta + V' (\theta) =
 \partial_\mu j_\mu^B,
\label{ddot-theta}
\ee
where ${ V(\theta) \approx m^2 \eta^2\left[-1+ (\theta-\pi)^2 \right]}$
and ${ j_\mu^B = \bar\psi \gamma_\mu \psi}$.
{Initially ${\theta}$ is uniform in ${ [0,2\pi]}$
and after inflation it started to oscillate around minimum.}

The second necessary equation is that for the quantum baryonic Dirac field:
\be
\left(i\partial+m  \right)\psi = -g\eta l+ 
(\partial_\mu\theta) \gamma_\mu \psi
\label{d-psi}
\ee
The solution to this equation can be found in one-loop approximation for 
${\psi (\theta)}$ in external classical field ${\theta}$. Then this solution, 
${ \psi^{\dagger} \psi = F (\theta)}$, should be substituted into eq. (\ref{ddot-theta}).
In this way a closed equation for $\theta (t)$ can be obtained. 
The solution oscillates with alternating baryonic number giving the  net result
for the baryon number density
\be
n_B \sim \eta^2 \Gamma_{\Delta B} (\Delta \theta)^3 .
\label{n_B-theta}
\ee

{\it The SUSY baryonic condensate scenario } will be discussed in more detail here
because with simple modification it allows for creation of astronomically significant
antimatter~\cite{ad-js}.

The basic features of this scenario are the following. 
SUSY predicts existence of scalars with non-zero baryonic number.
Such bosons may condense along {flat} directions of the potential:
\be
U_\lambda(\chi) = \lambda |\chi|^4 \left( 1- \cos 4\theta \right),
\label{U-lambda}
\ee
where ${ \chi = |\chi| \exp (i\theta)}$.
In SUSY models with high energy scale the baryonic number is naturally
non-conserved. It is reflected by the non-sphericity of potential (\ref{U-lambda}).
Due to infrared instability of massless (${ m\ll H}$) fields in de Sitter
space-time, $\chi$ can travel away from zero along the flat directions, 
$\theta = 0,\pi/2,\pi,3\pi/2 $.
We can also add a mass term to the potential:
\be
U_m( \chi ) = m^2 |\chi|^2 \left[ 1-\cos (2\theta+2\alpha) \right],
\label{U-m}
\ee
where ${ m=|m|e^\alpha}$. {If ${\alpha \neq 0}$, then C and CP are explicitly broken,
though it is not necessary for baryogenesis,

``Initially'' (as a result of inflation) ${\chi}$ was pushed away from origin and 
when inflation was over it started to evolve down to equilibrium point, ${\chi =0}$,
according to the equation of the Newtonian mechanics:
\be
\ddot \chi +3H\dot \chi +U' (\chi) = 0.
\label{ddot-chi}
\ee
The baryonic number of $\chi$:
\be
B_\chi =\dot\theta |\chi|^2
\label{B-chi}
\ee
is analogous to mechanical angular momentum. Using this mechanical analogy, and having 
the picture of potential $U(\chi)$ is easy to visualise the solution of the equation
of motion without explicitly solving it.

The baryonic number of $ \chi$ is  accumulated in its ``rotational'' motion,
induced by quantum fluctuations in orthogonal to valley direction.
When ${{\chi}}$ decays its 
baryonic charge is transferred to that of quarks through B-conserving processes.
If the mass term is absent or symmetric with respect to the phase rotation of $\chi$
this scenario leads to {globally charge symmetric universe.} 
The domain size ${l_B}$ is determined by the size of the region 
with a definite sign of ${ \dot\theta}$. Usually $l_B$ would be too small if no 
special efforts are done.

If ${m\neq 0}$, the angular momentum, B, is generated by a different 
direction of the mass valley at low ${\chi}$.
If CP-odd phase ${\alpha}$ is small but non-vanishing, both baryonic and 
antibaryonic regions are possible with dominance of one of them.
In this case {matter and antimatter domain may exist but globally ${B\neq 0}$}.

Now let us modify the model by adding general renormalizable 
coupling of $\chi$ to inflaton field ${\Phi}$:
\be
\lambda_\Phi |\chi|^2 \left( \Phi - \Phi_1 \right)^2,
\label{Phi-chi}
\ee
where $\Phi_1$ is the value of $\Phi$ which it passed during inflation, not too
long before its end. It is a free adjustable parameter.

Because of this coupling the gates to the valley would be open only for a short 
time when $\Phi$ was close to $\Phi_1$. So the probability for $\chi$ to 
reach a large value would be small. As a result we will have the following picture
of the universe. The bulk of space would have normal homogeneous baryon asymmetry, 
${ \beta = 6\cdot 10^{-10}}$, with small bubbles having large $\beta \sim 1$. In the 
simplest version of the scenario the high B regions should be almost
symmetric with respect to baryons and antibaryons.

The mass spectrum of such baryon rich bubbles is practically model independent
(it is determined  by inflation) and has simple log-normal form:
\be
\frac{dN}{dM} = C_0\, {\rm exp}\left[ -C_1 {\ln^2} \left(M/M_0\right)\right]
\label{dN-over-dM}
\ee
Such object could make primordial black holes, quasars, 
disperse clouds of antimatter, and unusual stars and anti-stars, all 
not too far from us in the Galaxy. Phenomenological implications of this
mechanism of antimatter creation and observational bounds are discussed in
ref.~\cite{anti-cb-ad}. If such mechanism is realized in nature the attempts
for search of cosmic antimatter have non-zero chances to be successful. 

It is worth noting that primordial nucleosynthesis in high B domains 
proceeded with large ratio ${ n_B/n_\gamma}$. So the outcome of light and heavier
element abundances could be much different from the predictions of the standard BBN.
Thus the regions in the sky with abnormal chemistry would be first candidates to 
search for cosmic antimatter through 100 MeV photons or through the positron
annihilation line.

\end{document}